\documentclass[12pt]{article}

\pdfoutput=1
\usepackage{jheppub}
\usepackage{graphicx,epsfig,amsmath,amssymb}
\usepackage{subcaption}
\usepackage{multirow}
\usepackage{slashed}

\newcommand{\als}{\ensuremath{\alpha_s}}

\newcommand{\mur}{\ensuremath{\mu_R}}
\newcommand{\amp}{\ensuremath{\mathcal{M}}}

\DeclareMathOperator{\vfin}{\mathcal{V}_{fin}}

\def\nn{\nonumber}

\def\eps{\epsilon}

\def\be{\begin{equation}}
\def\ee{\end{equation}}
\def\bea{\begin{eqnarray}}
\def\eea{\end{eqnarray}}
\def\nn{\nonumber}

\newcommand{\secdec}{\textsc{SecDec}{}}
\newcommand{\pysecdec}{py\secdec}
\newcommand{\gosam}{\textsc{GoSam}{}}
\newcommand{\powhegbox}{\texttt{POWHEG-BOX-V2}{}}

\newcommand{\mgg}{m_{\gamma\gamma}}
\newcommand{\mt}{m_t}

\title{Photon pair production in gluon fusion: Top quark
  effects at NLO with threshold matching}
\author[a]{Long Chen,}
\author[a]{Gudrun Heinrich,}
\author[a]{Stephan Jahn,}
\author[b]{Stephen P.~Jones,}
\author[c]{Matthias Kerner,}
\author[d]{Johannes Schlenk,}
\author[e]{Hiroshi Yokoya}

\affiliation[a]{Max Planck Institute for Physics, F\"ohringer Ring 6, 80805 M\"unchen, Germany}
\affiliation[b]{Theoretical Physics Department, CERN, Geneva, Switzerland}
\affiliation[c]{Physik-Institut, Universit{\"a}t Z{\"u}rich, Winterthurerstrasse 190, 8057 Z{\"u}rich, Switzerland}
\affiliation[d]{Theory Group LTP, Paul Scherrer Institut, CH-5232 Villigen PSI, Switzerland}
\affiliation[e]{Quantum Universe Center, KIAS, Seoul 02455, Korea}

\emailAdd{longchen@mpp.mpg.de}
\emailAdd{gudrun@mpp.mpg.de}
\emailAdd{sjahn@mpp.mpg.de}
\emailAdd{s.jones@cern.ch}
\emailAdd{mkerner@physik.uzh.ch}
\emailAdd{johannes.schlenk@psi.ch}
\emailAdd{hyokoya@kias.re.kr}

\preprint{{\small MPP-2019-236\\
    \hphantom{.}\hfill ZU-TH 48/19\\
    \hphantom{.}\hfill CERN-TH-2019-195\\
    \hphantom{.}\hfill PSI-PR-19-24}}

\abstract{
 We present a calculation of the NLO QCD corrections to the loop-induced production of a photon pair through gluon fusion, including massive top quarks at two loops,
where the two-loop integrals are calculated numerically.
Matching the fixed-order NLO results to a threshold expansion, we obtain accurate results around the top quark pair production threshold.
We analyse how the top quark threshold corrections affect distributions of the photon pair invariant mass and comment on the possibility of determining
the top quark mass from precision measurements of the diphoton invariant mass spectrum.
}

\keywords{LHC, two-loop computations, QCD/NRQCD phenomenology, direct photon production, top quark mass, top-quark threshold}

\begin{document}

\maketitle

\section{Introduction}

The production of pairs of photons in hadronic collisions has
attracted interest from both the experimental and the theory side
for several decades. 
Most prominently, the diphoton final state served as one of the key discovery
channels for the Higgs boson~\cite{Chatrchyan:2014fsa,Aaboud:2017vol}, which can decay into two photons. 
As a very clean experimental channel, it is also well suited for precision
studies of the Standard Model (SM) and in particular the Higgs sector. 
For example, there is the possibility to constrain the Higgs boson width from
interference effects of the continuum $gg\to \gamma\gamma$ spectrum with the
signal $gg\to H\to \gamma\gamma$~\cite{Dicus:1987fk,Dixon:2003yb,Martin:2012xc,deFlorian:2013psa,Martin:2013ula,Dixon:2013haa,Campbell:2017rke,Cieri:2017kpq}.
Furthermore, various New Physics models predict the production of photon
pairs, where the study of angular correlations between the decay
photons can provide information about the spin of the underlying
resonances~\cite{Aaboud:2016tru,Sirunyan:2018wnk}.

Another interesting aspect of diphoton production is the possibility of measuring the top quark mass via the top quark pair production threshold effects manifest in the diphoton invariant mass spectrum~\cite{Jain:2016kai,Kawabata:2016aya}.
While current LHC measurements~\cite{Aaboud:2017vol,Chatrchyan:2014fsa} are not yet able to provide the necessary statistics for such a threshold scan, the feasibility at the High-Luminosity LHC, and even more so at a future 100~TeV collider, is worth investigating.

Direct diphoton production\footnote{We denote by ``direct photons'' the photons produced directly in the hard scattering process, 
as opposed to photons originating from a hadron fragmentation process.} in hadronic collisions occurs via the leading order (LO) 
$\als^0$ process $q\bar{q}\to \gamma\gamma$.
The next-to-leading order (NLO) QCD corrections to this process, including fragmentation
contributions at NLO, were  implemented in the public program {\tt Diphox}~\cite{Binoth:1999qq}.

The loop induced $gg\to\gamma\gamma$ process enters as a next-to-next-to-leading order (NNLO) QCD (order $\als^2$)
correction to the $pp\to\gamma\gamma$ cross section.
The process  $gg\to\gamma\gamma$ has been calculated at LO including both massless and massive quark loops in Ref.~\cite{Dicus:1987fk} 
and is included in {\tt Diphox} at one loop for massless quark loops. 
Even though the $gg\to\gamma\gamma$ contribution is a higher-order correction to the total $pp\to \gamma\gamma$ cross section, 
its contribution is similar in size to the LO result at the LHC, due to the large gluon luminosity.
A calculation that includes also the effects of transverse-momentum resummation to direct photon production is
implemented in the program {\tt ResBos}~\cite{Balazs:2007hr}.

NLO QCD corrections to the gluon-fusion channel with massless quarks,
i.e. ${\cal O}(\als^3)$ corrections, have been first calculated in Refs.~\cite{Bern:2001df,Bern:2002jx} and implemented in the code 2$\gamma${\tt MC}~\cite{Bern:2002jx} as well as in MCFM~\cite{Campbell:2011bn}. 
Very recently, the NLO QCD corrections to the gluon-fusion channel including massive top quark loops have become available~\cite{Maltoni:2018zvp},
where the master integrals have been calculated numerically based on the numerical solution of differential equations~\cite{Czakon:2008zk, Mandal:2018cdj}.
Analytic results for the planar two-loop box integrals with massive top quarks have been presented in Ref.~\cite{Caron-Huot:2014lda,Becchetti:2017abb}. Regarding the non-planar contributions,  3-point topologies containing elliptic integrals have been calculated in Ref.~\cite{vonManteuffel:2017hms,Broedel:2019hyg}. Other 3-point topologies have been calculated earlier in the context of Higgs production and decay~\cite{Aglietti:2006tp,Anastasiou:2006hc}.

The NNLO QCD corrections to the process $pp\to \gamma\gamma$ were first calculated in Ref.~\cite{Catani:2011qz}, including the $gg\to\gamma\gamma$ contribution at order $\als^2$ with massless quark loops. For a phenomenological study see also Ref.~\cite{Catani:2018krb}.
The NNLO QCD corrections to  $pp\to \gamma\gamma$ have also been calculated and implemented in MCFM in Ref.~\cite{Campbell:2016yrh}, supplemented by the $gg$ initiated loops proportional to $n_f$ at LO and NLO  for five massless quark flavours, and at LO for massive top quark loops.
Diphoton production at NNLO with massless quarks is also available in {\sc Matrix}~\cite{Grazzini:2017mhc}.

The aim of this paper is twofold. Firstly, we provide an independent calculation of the QCD corrections to the process $gg\to\gamma\gamma$ including massive top quark loops, confirming the results of Ref.~\cite{Maltoni:2018zvp} for the central scale choice.
Secondly, we combine our results with threshold resummation as advocated in Ref.~\cite{Kawabata:2016aya}, such that the top quark pair production threshold region in the diphoton invariant mass spectrum can be predicted with high accuracy. The calculation can thus serve as a starting point for investigating the possibility of a top quark mass measurement from the diphoton invariant mass spectrum.

This work is structured as follows. In Section \ref{sec:calculation} we describe our calculation of the NLO corrections including both massless and massive fermion loops. Section \ref{sec:threshold} contains a description of our treatment of the top quark pair production threshold region. In Section \ref{sec:results} we present our numerical results. Finally, in Section \ref{sec:conclusions} we summarise and present an outlook on the possibility of measuring the top quark mass from the diphoton spectrum.

\section{Building blocks of the fixed order calculation}
\label{sec:calculation}


We consider the following scattering process,
\begin{align}\label{kinematicassignment}
g(p_1, \lambda_1, a_1) + g(p_2, \lambda_2, a_2) \to \gamma(p_3,\lambda_3) + \gamma(p_4,\lambda_4) ,
\end{align}
with on-shell conditions $p_j^2 = 0, j=1,...,4$.  
The helicities $\lambda_i$ of the external particles are defined by taking the momenta of the gluons 
$p_1$ and $p_2$ (with colour indices $a_1$ and $a_2$, respectively) 
as incoming and the momenta of the photons $p_3$ and $p_4$ as outgoing.
The Mandelstam invariants associated with
eq.~(\ref{kinematicassignment}) are defined by
\begin{align} \label{EQ:kinematicinvariants}
&s = \left(p_1 + p_2 \right)^2,&
&t = \left(p_2 - p_3 \right)^2,&
&u = \left(p_1 - p_3 \right)^2.
\end{align}

\subsection{Calculation of the virtual amplitudes}

\subsection*{Projection operators}

We define the tensor amplitude $\amp_{\mu_1\mu_2\mu_3\mu_4}$ by
extracting the polarisation vectors from the amplitude $\amp$,
\begin{equation} \label{eq:ggyyamplitudes}
    \amp{} =  \varepsilon_{\lambda_1}^{\mu_1}(p_1)\,\varepsilon_{\lambda_2}^{\mu_2}(p_2)\,\varepsilon_{\lambda_3}^{\mu_3,\star}(p_3)\,\varepsilon_{\lambda_4}^{\mu_4,\star}(p_4)\,\amp_{\mu_1\mu_2\mu_3\mu_4}(p_1,p_2,p_3,p_4),
\end{equation}
where the $\varepsilon_{\lambda_i}^{\mu_i}$ denote the polarisation vectors.
The amplitude is computed through projection onto a set of Lorentz
structures related to linear polarisation states of the external
massless bosons. An appropriate set of $D$-dimensional
projection operators is constructed following the approach proposed
in Ref.~\cite{Chen:2019wyb}, which has been applied recently in the
calculation of Ref.~\cite{Ahmed:2019udm}, and which we will summarise
briefly in the following.

A physical polarisation vector $\varepsilon(p)$ of a massless vector
boson with (on-shell) momentum $p$ fulfils 
the transversality and (imposed) normalisation conditions, 
\begin{equation}
    \varepsilon(p) \cdot p = 0, \quad \varepsilon(p) \cdot \varepsilon(p) = -1.
    \label{eq:ggyy:polvec_property}
\end{equation}
These conditions fix two components of the polarisation vectors in four space-time dimensions.
Now we construct explicitly a basis of the space of polarisation states defined by~\eqref{eq:ggyy:polvec_property} 
for the external massless vector bosons.
First, we introduce a polarisation basis vector $\varepsilon_X$, valid for both intial-state gluons, 
which can be written in terms of the linearly independent momenta of the process
\begin{equation}
    \varepsilon_X^\mu = c_1^X\,p_1^\mu + c_2^X\,p_2^\mu + c_3^X\,p_3^\mu\;,
\end{equation}
where the Lorentz invariant coefficients $c_i^X$ are determined by the system of equations
\begin{equation}
    \varepsilon_X \cdot p_1 = 0, \qquad
    \varepsilon_X \cdot p_2 = 0, \qquad
    \varepsilon_X \cdot \varepsilon_X = -1.
\end{equation}
Note that the conditions above constitute a gauge choice in which the reference momentum of either incoming gluon is set to be the momentum of the other gluon.
A polarisation vector $\varepsilon_T$ for both outgoing photons can be constructed analogously: 
\begin{equation}
    \varepsilon_T \cdot p_3 = 0, \qquad
    \varepsilon_T \cdot p_4 = 0, \qquad
    \varepsilon_T \cdot \varepsilon_T = -1.
\end{equation}
A third basis vector $\varepsilon_Y$, pointing out of the scattering plane,
is needed to span the space of all possible polarisation vectors for this process:
\begin{equation}
    \varepsilon_Y \cdot p_i = 0,\qquad i\in \{1,\ldots,4\}\;.
    \label{eq:ggyy:definiing_property_eps_Y}
\end{equation}
In four dimensions, such a vector can be constructed using the Levi-Civita tensor:
\begin{equation}
    \varepsilon_Y^\mu = \varepsilon^{\mu\nu\rho\sigma} {p_1}_\nu \, {p_2}_\rho \, {p_3}_\sigma.
\end{equation}
Since we consider only QCD corrections to a QED process, neither $\gamma{}_5$ nor Levi-Civita tensors are
introduced by the relevant Feynman rules. Consequently, a completely $D$-dimensional tensor decomposition 
of this scattering amplitude can be expressed solely in terms of metric tensors and external momenta. 
Therefore, a contraction of the tensor amplitude with an odd number of $\varepsilon_Y$ evaluates to zero.
A product of two Levi-Civita tensors, however, can be rewritten in terms of metric tensors using
\begin{equation}
  \varepsilon^{\mu\nu\rho\sigma}\, \varepsilon{}_{\alpha\beta\kappa\lambda} =
  \det \left( \begin{array}{cccc}
     \delta_{\alpha}^{\mu}   & \delta_{\beta}^{\mu} &
     \delta_{\kappa}^{\mu}   & \delta_{\lambda}^{\mu} \\
     \delta_{\alpha}^{\nu}   & \delta_{\beta}^{\nu} &
     \delta_{\kappa}^{\nu}   & \delta_{\lambda}^{\nu} \\
     \delta_{\alpha}^{\rho}  & \delta_{\beta}^{\rho} &
     \delta_{\kappa}^{\rho}  & \delta_{\lambda}^{\rho} \\
     \delta_{\alpha}^{\sigma}& \delta_{\beta}^{\sigma} &
     \delta_{\kappa}^{\sigma}& \delta_{\lambda}^{\sigma}
   \end{array} \right),
   \label{eq:LeviCivita_combination}
\end{equation}
which has a straightforward $D$-dimensional continuation.
For a detailed discussion of the subtleties related to the manipulation of Levi-Civita tensors in the construction of 
projectors for more general cases we refer to Ref.~\cite{Chen:2019wyb}.
~\\

Applied to the scattering process \eqref{kinematicassignment}, this
construction leads to eight projectors  
\begin{equation} \label{eq:LPprojectors}
    \varepsilon_{[X,Y]}^\mu \varepsilon_{[X,Y]}^\nu \varepsilon_{[T,Y]}^\rho \varepsilon_{[T,Y]}^\sigma,
\end{equation}
where the square bracket $[\cdot{},\cdot{}]$ means either entry and where only the combinations containing an even number of $\varepsilon_Y$ are considered.
Let us emphasize again that, in order to avoid possible ambiguities in the application of 
these projectors, all pairs of Levi-Civita tensors are replaced
according to the contraction rule~\eqref{eq:LeviCivita_combination}
{\em before} being used for the projection of the amplitude.
Then the aforementioned projectors are expressed solely in terms of external momenta and metric tensors 
whose open Lorentz indices are all set to be $D$-dimensional.

The usual helicity amplitudes can be constructed as circular polarisation states from the linear ones using the relations 
\begin{align}
  \begin{split}
    \varepsilon_{\pm}(p_1)^\mu &= \frac{1}{\sqrt{2}} \left( \varepsilon_X^\mu \pm i \varepsilon_Y^\mu \right), \\
    \varepsilon_{\pm}(p_2)^\nu &= \frac{1}{\sqrt{2}} \left( \varepsilon_X^\nu \mp i \varepsilon_Y^\nu \right), \\
    \varepsilon_{\pm}(p_3)^\rho &= \frac{1}{\sqrt{2}} \left( \varepsilon_T^\rho \pm i \varepsilon_Y^\rho \right), \\
    \varepsilon_{\pm}(p_4)^\sigma &= \frac{1}{\sqrt{2}} \left( \varepsilon_T^\sigma \mp i \varepsilon_Y^\sigma \right).
  \end{split}
\end{align}

Analytic results for the LO amplitudes of \eqref{kinematicassignment} were obtained 
quite some time ago  in Refs.~\cite{Karplus:1950zz,Bern:2001dg,Binoth:2002xg} for massless quark loop contributions and in Refs.~\cite{Bern:1995db,Bernicot:2008th} with massive quark loop contributions.
With the linear polarisation projectors defined in
\eqref{eq:LPprojectors}, we re-computed these LO amplitudes
analytically, with both  massless and massive quark loops.
These expressions were implemented in our computational setup for the
NLO QCD corrections to the considered process, which
we describe below.

\subsection*{UV renormalisation}

The bare scattering amplitudes of the process \eqref{kinematicassignment},
denoted by $\hat{\amp}$, beyond LO  contain poles in the dimensional regulator $\epsilon \equiv (4-D)/2$ arising from ultraviolet (UV) as well as soft and collinear (IR) regions of the loop momenta. 
In our computation, we renormalise these UV divergences using the
$\overline{\text{MS}}$ scheme, 
except for the top quark mass which is  renormalised on-shell. 

The bare virtual amplitude $\hat{\amp}$ is a function of the bare QCD coupling $\hat{\alpha}_s$ and the
bare top quark mass $\hat{m}_t$.
The UV renormalisation of $\hat{\amp}$ is achieved by the replacement 
\begin{align}
\hat{\alpha}_s\,\hat{\mu}^{2\eps}\,S_\eps = \alpha_s\, \mu_R^{2\eps}\, Z_a\; , \;
\hat{m}_t = m_t\,Z_m,
\end{align}
and by renormalising the gluon wave function.
Here,  $S_{\eps} = \left(4 \pi \right)^{\eps} e^{-\eps \gamma_E}$, with $\gamma_E$ the Euler constant. The strong coupling is given by $\alpha_s = g_s^2/(4 \pi)$ and $\hat{\mu}$ is an auxiliary
mass-dimensionful parameter introduced in dimensional regularisation
to keep the coupling constants dimensionless. The usual renormalisation scale is denoted $\mu_R$, and we will use $\hat{\mu}=\mu_R$ in the following.

Both the bare virtual amplitudes and the UV renormalisation constants are expanded in $a_s \equiv \alpha_s(\mur)/(4\pi)$.
We may write the renormalisation constants as
\begin{equation}
Z_{i} = 1 + a_s \,\delta Z_{i}  + O(a_s^2), \qquad i=a,A,m.
\end{equation}
Under the $\overline{\text{MS}}$ scheme for $\alpha_s$ with $n_f$ massless quark flavours and top-quark loops renormalised on-shell, the renormalisation constants needed in our computation read
\begin{align} \label{ZuvCoeffs}
\delta Z_{a} &= -\frac{1}{\eps}\,\beta_0 + \left(\frac{\mur^2}{m_t^2}\right)^{\eps}\frac{4}{3\eps}\,T_R, \nn \\
\delta Z_A &= \left(\frac{\mur^2}{m_t^2}\right)^{\eps}\,\left(-\frac{4}{3\eps}\,T_R\right), \nn \\
\delta Z_m &= \left(\frac{\mur^2}{m_t^2}\right)^{\eps}\,C_F\,\left(-\frac{3}{\eps}-4\right),
\intertext{with}
\beta_0 &= \frac{11}{3}C_A-\frac{4}{3}\,T_R \,n_f.
\end{align}
We write the scattering amplitude for the
process $gg\to \gamma\gamma$, up to second order in $a_s$, in the following form 
\begin{align}
\hat{\amp} = & \hat{a}_s \hat{\mathcal{M}}_B(\hat{m}_t) + 
\hat{a}_s^2 \hat{\mathcal{M}}_V(\hat{m}_t) + \mathcal{O}(\hat{a}^3_s) \nonumber\\
= & a_s\, \mathcal{M}_{B,\mathrm{ren}}(m_t) + a_s^2\, \mathcal{M}_{V,\mathrm{ren}}(m_t) + \mathcal{O}(a^3_s), 
\end{align}
where
\begin{align}
\mathcal{M}_{B,\mathrm{ren}}(m_t)  = & S_\eps^{-1} \hat{\mathcal{M}}_B(\hat{m}_t) \nonumber \\
\mathcal{M}_{V,\mathrm{ren}}(m_t)  = & S_\eps^{-2}  \hat{\mathcal{M}}_V(\hat{m}_t) - \frac{\beta_0}{\eps} S_\eps^{-1} \hat{\mathcal{M}}_B(\hat{m}_t) + \delta Z_m \ \hat{\mathcal{M}}_{CT}(\hat{m}_t).
\end{align}
Here, $\mathcal{M}_{B,\mathrm{ren}}(m_t)$ and $\mathcal{M}_{V,\mathrm{ren}}(m_t)$ are the one-loop and UV renormalised two-loop amplitudes, respectively, with the Born kinematics given in \eqref{kinematicassignment}. The mass counter-term amplitude $\hat{\mathcal{M}}_{CT}(\hat{m}_t)$ is obtained by inserting a mass counter-term into the heavy quark propagators
\begin{equation}
\Pi^{\delta_m}_{ab}(p) = \frac{i \delta_{ac} }{\slashed{p}-m} (-i \delta Z_m) \frac{i \delta_{cb} }{\slashed{p}-m},
\end{equation}
where $a,b,c$ are colour indices in the fundamental representation. The mass counter-term can also be obtained by taking the derivative of the one-loop amplitude with respect to $\hat{m}_t$.

\subsection*{Definition of the IR-subtracted virtual part}

The UV renormalised virtual amplitude $\mathcal{M}_{V,\mathrm{ren}}$
still contains divergences arising from soft and collinear
configurations of the loop momenta, which appear as poles in the dimensional regulator.
We employ the FKS subtraction approach~\cite{Frixione:1995ms} to deal with the intermediate IR divergences,
as implemented in the \powhegbox~framework~\cite{Nason:2004rx,Frixione:2007vw,Alioli:2010xd}.

For the process $gg \rightarrow \gamma \gamma$, the corresponding integrated subtraction operator is given by
\begin{equation}
I_1(\mur^2,s) = \frac{S_\epsilon^{-1}}{\Gamma(1-\epsilon)} \left[ \frac{2 C_A}{\epsilon^2} + \frac{2 \beta_0}{\epsilon} + \frac{2 C_A}{\epsilon} \ln \left( \frac{\mur^2}{s} \right) \right].
\end{equation}
To second order in $a_s$ the UV renormalised and IR subtracted virtual amplitude is given by
\begin{align}
\mathcal{M}_B=& \mathcal{M}_{B,\mathrm{ren}}, \nonumber\\
\mathcal{M}_V =& \mathcal{M}_{V,\mathrm{ren}}+ I_1(\mur^2,s)\ \mathcal{M}_{B,\mathrm{ren}}.
\end{align}
Note that the LO amplitude $\mathcal{M}_{B,\mathrm{ren}}$ needs to be computed 
to $\mathcal{O}(\epsilon^2)$ as it is multiplied by coefficients
containing $1/\eps^2$ poles.

In practice,  we need to supply only the finite part of the
born-virtual interference, under a specific
definition~\cite{Alioli:2010xd} in order to combine it with the
FKS-subtracted real radiation generated within the
\gosam/\powhegbox~framework. Explicitly, we compute
\begin{align}
{\cal V}_{\rm{fin}}(\mu_R) =  a_s^2(\mu_R)\,\mathrm{Re} \left[ \mathcal{M}_V
  \mathcal{M}_B^\dagger \right].
  \label{eq:def_Vfin}
\end{align}

The renormalisation scale dependence of $\vfin$ can be derived from the above definitions, it is given by
\begin{equation} \label{eq:VfinNote}
\vfin{}(\mu_R) = 
\vfin{}(\mu_0) \left(\frac{a_s(\mu_R)}{a_s(\mu_0)}\right)^2
+ \left[ C_A \log{}^2 \left( \frac{\mu_0^2}{s} \right) - C_A \log{}^2 \left( \frac{\mu_R^2}{s} \right) \right]
a^2_s(\mu_R)\left|\mathcal{M}_B\right|^2,
\end{equation}
where $\mu_0$ stands for an arbitrarily chosen initial renormalisation scale.


\subsection*{Evaluation of the virtual amplitude}

\begin{figure}[htb]
\centering
\begin{subfigure}[b]{0.4\textwidth}
  \centering
  \includegraphics[width=0.8\textwidth]{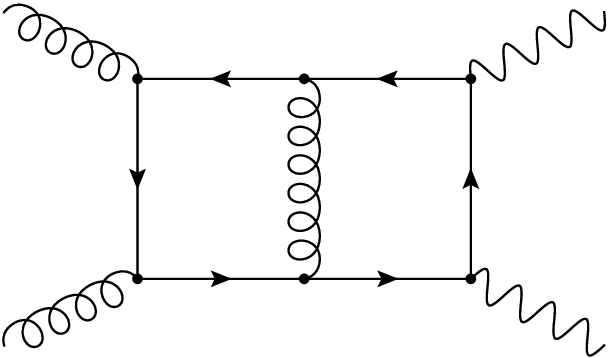} 
\end{subfigure}
\begin{subfigure}[b]{0.4\textwidth}
  \centering
  \includegraphics[width=0.8\textwidth]{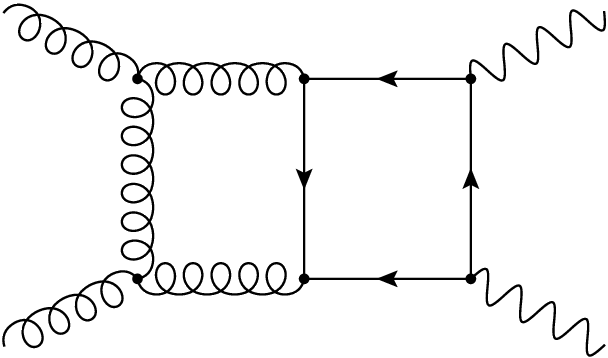} 
\end{subfigure}
\vspace{1em}\\
\begin{subfigure}[b]{0.4\textwidth}
  \centering
  \includegraphics[width=0.8\textwidth]{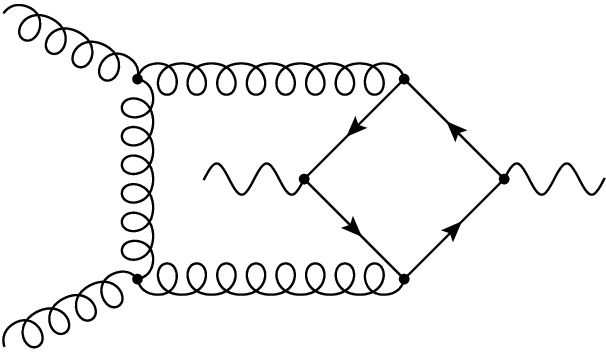} 
\end{subfigure}
\begin{subfigure}[b]{0.4\textwidth}
  \centering
  \includegraphics[width=0.8\textwidth]{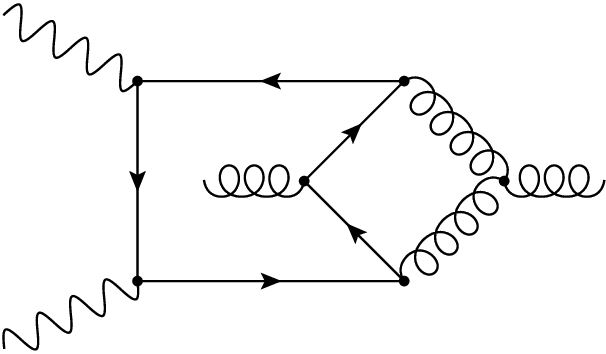} 
\end{subfigure}
\caption{Examples of diagrams contributing to the virtual corrections.}
\label{fig:2l_diagrams}
\end{figure}

For the two-loop QCD diagrams contributing to our scattering process there is a complete separation of quark flavours due to the colour algebra and Furry's theorem. 
Consequently we have $n_f + 1$ sets of two-loop diagrams which can be treated separated from each other.
The two-loop amplitude has been obtained with the multi-loop extension of the program~\gosam{}~\cite{Jones:2016bci} where {\sc Reduze}\,2~\cite{vonManteuffel:2012np} is employed for the reduction to master integrals. 
In particular, each of the linearly polarised amplitudes projected out using \eqref{eq:LPprojectors} 
is eventually expressed as a linear combination of 39 massless 
integrals and 171 integrals that depend on the top quark mass, distributed into three integral families.
All massless two-loop master integrals involved are known analytically~\cite{Bern:2001df,Binoth:2002xg,Argeri:2014qva},
and we have implemented the analytic expressions into our code.   
Regarding the two-loop massive integrals which are not yet fully known analytically,
we first rotate to an integral basis consisting partly of quasi-finite loop integrals~\cite{vonManteuffel:2014qoa}. Our integral basis is chosen such that the second Symanzik polynomial, $\mathcal{F}$, appearing in the Feynman representation of each of the integrals is raised to a power, $n$, where $|n| \le 1$ in the limit $\epsilon \rightarrow 0$. This choice improves the numerical stability of our calculation near to the $t \bar{t}$ threshold, where the $\mathcal{F}$ polynomial can vanish. The integrals are then evaluated numerically using \pysecdec~\cite{Borowka:2017idc,Borowka:2018goh}.
Examples of contributing two-loop Feynman diagrams are shown in Figure~\ref{fig:2l_diagrams}.

The phase-space integration of $\vfin$ is achieved by reweighting unweighted Born events.
The accuracy goal imposed on the numerical evaluation of the virtual
two-loop amplitudes in the linear polarisation basis in \pysecdec{} is 1
per-mille on both the relative and the absolute error.
We have collected 6898 phase space points out of
which 862 points fall into the diphoton invariant mass
window $\mgg \in \left[330,\, 360\right]$ GeV.
We have also calculated a further 2578 phase space
points restricted to the threshold region.

\subsection{Computation of the real radiation contributions}

\begin{figure}[htb]
\centering
\vspace{1em}
\begin{subfigure}[b]{0.3\textwidth}
  \centering
  \includegraphics[width=0.8\textwidth]{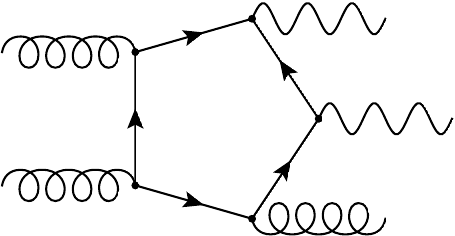}
\end{subfigure}
\begin{subfigure}[b]{0.3\textwidth}
  \centering
  \includegraphics[width=0.8\textwidth]{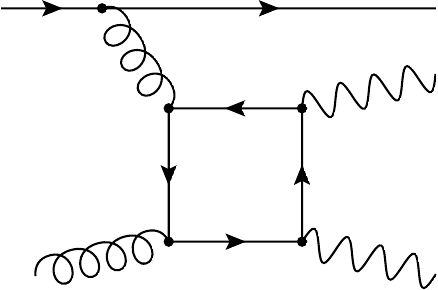}
\end{subfigure}
\begin{subfigure}[b]{0.3\textwidth}
  \centering
  \includegraphics[width=0.8\textwidth]{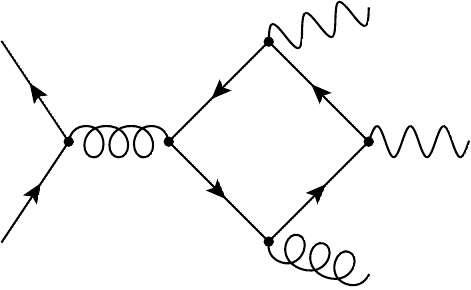}
\end{subfigure}
\caption{Examples of diagrams contributing to the real radiation part.}
\label{fig:real_rad_diagrams}
\end{figure}

The real radiation matrix elements are calculated using the interface~\cite{Luisoni:2013cuh}
between \gosam~\cite{Cullen:2011ac,Cullen:2014yla} and the \powhegbox~\cite{Nason:2004rx,Frixione:2007vw,Alioli:2010xd}, modified accordingly to compute the real radiation corrections to loop-induced Born amplitudes. 
Only real radiation contributions which contain a closed quark loop at the amplitude level are included.
We also include the $q\bar{q}$ initiated diagrams which contain a closed quark loop, even though their contribution is numerically very small.
Examples of Feynman diagrams contributing to the real radiation amplitude are shown in Figure~\ref{fig:real_rad_diagrams}.
The diagrams in which one of the photons is radiated off a closed fermion loop and the
other photon is radiated off an external quark line
vanish due to Furry's theorem. 

\section{Treatment of the threshold region}
\label{sec:threshold}


When the partonic centre-of-mass energy is close to the threshold for
the production of a $t\bar{t}$ pair, the top quarks are produced with
a non-relativistic velocity
such that Coulomb interactions between the top
quarks can play a significant role.  In the case of 
the top-loop induced contribution to diphoton production, the Coulomb singularity
appears in the form of a logarithmic dependence on the velocity first at two-loop order, due to the exchange of a soft gluon
between the top quarks in the loop.
To overcome this issue and correctly describe the threshold, we employ the so-called non-relativistic QCD (NRQCD)~\cite{Caswell:1985ui,Bodwin:1994jh,Pineda:1997bj,Beneke:1997zp}, which is an effective field theory
designed to describe non-relativistic heavy quark-antiquark systems in the threshold region.

\subsection{NRQCD amplitude}
\label{sec:pnrqcd_amp}

To the order which we consider here, the amplitude can be expressed
as a coherent sum of light quark loop contributions and the top quark loop contributions,
\begin{align}
\amp{}(p_i,\lambda_i,a_1,a_2) 
=  8 \alpha_e \alpha_s\ T_R\,\delta^{a_1a_2}
\left[ \left(\sum_{q} Q_q^2\right) \mathrm{M}_{q}(s,t) 
+ Q_t^2 \,\mathrm{M}_{t}(s,t)\right],
\label{eq:ampdef}
\end{align}
where $\alpha_e = e^2/(4 \pi)$ and $Q_{q}$ denotes the electric charge of quark $q$. 
In our computation, the NRQCD expansion of the amplitude $\mathrm{M}_{t}$ near the $t\bar{t}$ 
threshold is performed according to the formalism explained in more detail in Refs.~\cite{Melnikov:1994jb,Kawabata:2016aya}.
Near the production threshold of an intermediate $t\bar{t}$ pair, $m_{\gamma\gamma} \simeq 2 m_t$, we define
\begin{align}
&E \equiv m_{\gamma\gamma}-2m_t,&
&\beta \equiv \sqrt{1-4m_t^2/\mgg^2+i\delta},&
\end{align}
and the scattering angle is given by
\begin{equation}
\cos\theta = 1 +
t\,(1-\beta^2) / (2\,\mt^2).
\end{equation}
Close to threshold, the amplitude
$\mathrm{M}_{t}$ can be parametrised as~\cite{Melnikov:1994jb,Kawabata:2016aya}
\begin{align}
{\mathrm M}_{t}^{\rm NR} = {\mathcal A}_t(\theta) +
{\mathcal B}_t(\beta)\, G(\vec0;{\mathcal E}) + \mathcal{O}(\beta^3),
\label{eq:thr}
\end{align}
where ${\mathcal E}=E+i\Gamma_t$ includes the top-quark decay width 
$\Gamma_t$\footnote{It has been shown in Ref.~\cite{Melnikov:1993np} that in the non-relativistic limit the top width can be consistently included by calculating the cross section for stable top quarks supplemented by such a replacement up to next-to-leading-order according to the NRQCD power counting.}. Note that the P-wave contribution ${\mathcal B}_{t,P}(\beta)\, G_P(\vec0;{\mathcal E})$ starts at $\mathcal{O}(\beta^3)$.
In this parametrisation, the amplitude ${\mathrm M}_{t}^{\rm NR}$ is split into two parts: ${\mathcal
  B}_t(\beta)\, G(\vec0;{\mathcal E})$, which contains the
$t\bar{t}$ bound state effects, and $ {\mathcal A}_t(\theta)$, which
does not. The  term ${\mathcal B}_t(\beta)\, G(\vec0;{\mathcal
  E})$ contains the effects from resumming the non-relativistic static
potential interactions, where the Green's function $G(\vec0;{\mathcal
  E})$ is obtained by solving the non-relativistic Schr\"odinger
equation describing a colour-singlet $t\bar{t}$ bound state:
\begin{equation}
\left( -\frac{\nabla^2}{m_t} + V(r) - {\mathcal E} \right) G(\vec{r};{\mathcal E}) =
\delta (\vec{r}),
\label{eq:schroedinger}
\end{equation} 
with the QCD static potential~\cite{Fischler:1977yf,Billoire:1979ih}
\begin{equation}
V(r) = -C_F \frac{\als(\mu)}{r} 
\Bigg( 
1 + \frac{\als(\mu)}{4 \pi} 
\left(2 \beta_0 \left( \mathrm{ln}(\mu \, r) + \gamma_E \right) 
+ \frac{31}{9}C_A - \frac{10}{9} n_f \right)
\Bigg)
+ {\mathcal O}(\als^3)\, .
\end{equation}
The mass $m_t$ appearing in \eqref{eq:schroedinger} is the
pole mass of the top quark. 
$G(\vec0;{\mathcal E})$ is the $r \rightarrow 0$ limit of the Green's
function $G(\vec{r};{\mathcal E})$.
The real part of the NLO Green's function at $r=0$ is divergent and therefore has to be renormalised. We adopt the $\overline{\rm MS}$ scheme, thus introducing a scale $\mu$ into the renormalised Green's function~\cite{Beneke:1998rk,Hoang:1998xf, Beneke:1999ff,Hoang:2001mm}. 
The coefficient ${\mathcal B}_t(\beta)$ can be obtained from the Wilson coefficients of the   
$gg t\bar{t}$ and $\gamma \gamma t\bar{t}$ operators~\cite{Kawabata:2016aya} in the NRQCD 
effective Lagrangian for the process $gg\to \gamma\gamma$.
The term ${\mathcal A}_t(\theta)$ encompasses the non-resonant corrections, resulting from quark loops 
with large virtuality which can be systematically computed order by order in $\alpha_s$.

Both ${\mathcal A}_t$ and  ${\mathcal B}_t$ can be expanded perturbatively in $\als$. 
For the process $gg\to \gamma\gamma$,
corrections to ${\mathcal B}_t$ have been calculated up to ${\cal
  O}(\als)$ and ${\cal O}(\beta^2)$ in Ref.~\cite{Kawabata:2016aya}, where
explicit expressions of ${\mathcal B}_t$ at the leading order for all
relevant helicity configurations can be found. Here we repeat for
completeness the expressions for the
$S$--wave  $t\bar{t}$ resonance we are considering. For the $S$--wave 
the ${\mathcal B}_t$ coefficients are independent of the scattering
angle.
We use the notation $ G(\beta) \equiv G(\vec0; E)$ and
\begin{align}
 {\cal M}_{t,\{\lambda_i\}}^{\rm NR} & = {\mathcal A}_{t,\{\lambda_i\}}(\theta) +  {\mathcal B}_{t,\{\lambda_i\}}(\beta)\,G(\beta) \nn\\
 &= {\cal M}_{t,\{\lambda_i\}}^{\rm NR, (0)}+\frac{\als}{\pi}\,{\cal M}_{t,\{\lambda_i\}}^{\rm NR, (1)}+{\cal O}(\als^2)\;.
\label{eq:expandM}
\end{align}
Note that an overall factor of $\als$ already has been extracted from the amplitude (see Eq.~(\ref{eq:ampdef})), 
such that the ${\cal O}(\als)$ term in the expression (\ref{eq:expandM}) contains the two-loop amplitude.
The NLO part of ${\cal M}_{t}^{\rm NR}$, denoted by ${\cal M}_{t}^{\rm NR, (1)}$, can be expanded as
\begin{align}
{\cal M}_{t}^{\rm NR, (1)} &=  A^{(1)}_t(\theta) + B^{(1)}_t(\beta) G^{(0)}(\beta) +B^{(0)}_t(\beta) G^{(1)}(\beta).
\label{eq:2loopexpansion}
\end{align}
The expression for $B^{(n)}_t$ can be further expanded in $\beta$,
\begin{equation}
B^{(n)}_t(\beta)=b^{(n)}+\beta^2\,\tilde{b}^{(n)}+{\cal O}(\beta^3),
\end{equation}
where~\cite{Kawabata:2016aya,Petrelli:1997ge,Hagiwara:2008df,Kiyo:2008bv} 
\begin{align}
b^{(0)}_{\{\lambda_i\}}&=-\frac{4\pi^2}{m_t^2}\,\lambda_1\lambda_3\,\delta_{\lambda_1\lambda_2}\delta_{\lambda_3\lambda_4},\nn\\
\tilde{b}^{(0)}_{\{\lambda_i\}}&=-\frac{16\pi^2}{3m_t^2}\,\lambda_1\lambda_3\,\delta_{\lambda_1\lambda_2}\delta_{\lambda_3\lambda_4},\nn\\
b^{(1)}_{\{\lambda_i\}}&=b^{(0)}_{\{\lambda_i\}}\,b_1,\;
\tilde{b}^{(1)}_{\{\lambda_i\}}=\tilde{b}^{(0)}_{\{\lambda_i\}}\,b_1,\nn\\
b_1 &= C_F \left( - 5 +
 \frac{\pi^2}{4} \right) + \frac{C_A}{2} \left( 1 + \frac{\pi^2}{12}
 \right) + \frac{\beta_0}{2} \ln{\left(\frac{\mu}{2m_t}\right)}. \label{eq:b1}
\end{align}
The expansion of the Green's function in $\alpha_s$ is given by
\begin{align}
G(\beta) =& G^{(0)}(\beta) + \frac{\alpha_s}{\pi} G^{(1)}(\beta,\mu) + \mathcal{O}(\als^2),
\end{align}
where~\cite{Hoang:2001mm,Hoang:2004tg}
\begin{align}
G^{(0)}(\beta)=& i\,\frac{m_t^2}{4\pi}(\beta+\beta^3) + \mathcal{O}(\beta^5),\\
G^{(1)}(\beta,\mu)=& \frac{m_t^2}{8}\,C_F\left(1-2\ln(-i\beta)+2\ln(\frac{\mu}{2m_t})+\beta^2\,[1-4\ln(-i\beta)+4\ln(\frac{\mu}{2m_t})]\right.\nonumber\\
&\left.+i\beta^3\frac{16}{3\pi}\,[2c_{us}+2\ln(-i\beta)-\ln(\frac{\mu}{2m_t})]\right) + \mathcal{O}(\beta^4),\\
c_{us}=&-\frac{7}{4}+\ln{2}\nonumber.
\end{align}

For ${\mathcal A}_t(\theta)$, we can make use of a partial-wave decomposition in terms of Wigner functions $d^{J}_{hh'}(\theta)$,
\begin{align}
 {\mathcal A}_{t,\{\lambda_i\}} (\theta) = \sum_{J=0}^{\infty}
 (2J+1){\mathcal A}^{J}_{t,\{\lambda_i\}} d^{J}_{hh'}(\theta),
 \label{eq:partialwaves}
\end{align}
where $h=-\lambda_1+\lambda_2$ and $h'=\lambda_3-\lambda_4$.

\subsection{NRQCD-improved calculation}
\label{sec:pnrqcd_Xsect}

\subsection*{Matched amplitude}

We would like to retain NRQCD resummation effects and, at the same time, keep the 
cross section accurate up to NLO in the fixed-order power counting.  
We define the ``NRQCD-matched'' amplitude as~\cite{Kawabata:2016aya}
\begin{align} \label{eq:match}
{\mathrm M}_{t}^{\rm match}  & \equiv {\mathrm M}_{t} + {\mathcal B}_{t} \, G(\vec0;{\mathcal E}) - {\mathrm M}_{\mathrm{OC}},
\end{align}
where the first term is the fixed-order amplitude, the second term describes the threshold according to NRQCD and the third term $ {\mathrm M}_{\mathrm{OC}}  \equiv {\mathcal B}_{t} \, G(\vec0;E)$ subtracts double counted contributions included in both the fixed-order amplitude and NRQCD contribution. 
The $ {\mathrm M}_{\mathrm{OC}}$ term in a fixed-order computation should be expanded to the same order as the fixed-order amplitude.

Expanding \eqref{eq:match} to next-to-leading order, we have
\begin{align}
{\mathrm M}_{t} & = {\mathrm M}_{t,B} + \frac{\alpha_s}{\pi} {\mathrm M}_{t,V} + \mathcal{O}(\alpha_s^2), \nn\\
{\mathrm M}_{\mathrm{OC}} & = {\mathrm M}_\mathrm{OC}^{(0)} + \frac{\alpha_s}{\pi}  {\mathrm M}_\mathrm{OC}^{(1)} + \mathcal{O}(\alpha_s^2),
\end{align}
with
\begin{align}
{\mathrm M}_{\mathrm{OC}}^{(0)} & = \mathcal{B}_t^{(0)} G^{(0)}(\vec0;E), \nn \\
{\mathrm M}_{\mathrm{OC}}^{(1)} & = \mathcal{B}_t^{(1)} G^{(0)}(\vec0;E) + \mathcal{B}_t^{(0)} G^{(1)}(\vec0;E).
\end{align}
Inserting into the matched amplitude we obtain,
\begin{align}
{\mathrm M}_{t}^{\rm match}  = \left[ {\mathcal B}_{t} \, G(\vec0;{\mathcal E}) + ({\mathrm M}_{t,B} - {\mathrm M}_{\mathrm{OC}}^{(0)}) \right] + \frac{\alpha_s}{\pi}  \left[  {\mathrm M}_{t,V} - {\mathrm M}_{\mathrm{OC}}^{(1)} \right] + \mathcal{O}(\alpha_s^2).
\end{align}
The NLO-matched cross section is obtained by squaring the matched amplitude and adding the corresponding real-radiation.
Upon squaring the matched amplitude we obtain,
\begin{align}
|{\mathrm M}_{t}^{\rm match}|^2 = & \left| {\mathcal B}_{t} \, G(\vec0;{\mathcal E}) + ({\mathrm M}_{t,B} - {\mathrm M}_{\mathrm{OC}}^{(0)}) \right|^2 \nonumber \\
& + \frac{\alpha_s}{\pi} \ 2  \mathrm{Re} \left[  {\mathrm M}_{t,B}^\dagger (M_{t,V} - {\mathrm M}_{\mathrm{OC}}^{(1)}) \right]  \\
& + \frac{\alpha_s}{\pi} \ 2  \mathrm{Re} \left[ ( {\mathcal B}_{t} \, G(\vec0;{\mathcal E}) - {\mathrm M}_{\mathrm{OC}}^{(0)} )^\dagger ({\mathrm M}_{t,V} - {\mathrm M}_{\mathrm{OC}}^{(1)}) \right] + \mathcal{O}(\alpha_s^2). \label{eq:match_amp_squared}
\end{align}
Expanding the $( {\mathcal B}_{t} \, G(\vec0;{\mathcal E}) - {\mathrm M}_{\mathrm{OC}}^{(0)} )$ term we note that the last line is formally of order $\alpha_s^2$ (i.e. beyond NLO accuracy) and we do not include it in our calculation. However, in the first line, we retain the full ${\mathcal B}_{t} \, G(\vec0;{\mathcal E})$ term, which describes the threshold behaviour. The fixed-order massless quark contribution can be included by replacing the top-quark only amplitude, $\mathrm{M}_t$, with the full amplitude and restoring overall factors extracted from the top-only amplitude. 

\subsection*{Matched cross section}

We define our NLO-matched cross section as follows 
\begin{eqnarray}
\sigma^\mathrm{match}_\mathrm{LO} &\equiv&
a_s^2(\mu_R) \int_{\tau_{min}}^{1} \mathrm{d} \tau \frac{\mathrm{d} \mathcal{L}_{gg}(\mu_F)}{\mathrm{d} \tau} \, \mathcal{N}_{gg}
\int \mathrm{d}\Phi_2 \, \Big| \mathcal{M}_{B} + 
\mathbf{c}\, \left( {\mathcal B}(\mu)\, G(\vec0;{\mathcal E}, \mu) 
- {\mathrm M}_{\rm{OC}}^{(0)} \right) \Big|^{\,2}, \nonumber\\
\nonumber\\
\sigma^{{\rm match}}_\mathrm{NLO} &\equiv& \sigma^\mathrm{match}_\mathrm{LO} \nonumber\\
&+& a_s^3(\mu_R)\, \int_{\tau_{min}}^{1} \mathrm{d} \tau \frac{\mathrm{d} \mathcal{L}_{gg}(\mu_F)}{\mathrm{d} \tau} \, \mathcal{N}_{gg} \,
\int \mathrm{d}\Phi_2 \, 2\,\mathrm{Re} \left[ \mathcal{M}_{B}^{\dagger} 
\left( {\cal M}_V(\mu_R) - \mathbf{c}\,{\mathrm M}_{\rm{OC}}^{(1)}(\mu) \right) \right] \nonumber\\
&+& a_s^3(\mu_R) \int_{\tau_{min}}^{1} \mathrm{d} \tau  \sum_{ij} \frac{\mathrm{d} \mathcal{L}_{ij}(\mu_F)}{\mathrm{d} \tau} \, 
\mathcal{N}_{ij}\, \int \mathrm{d}\Phi_3 \, \Big|{\cal M}_{R,[ij]}(\mu_R) \Big|^{\,2} 
+ \sigma_C \left( \mu_F, \mu_R \right), \nn \\
\label{eq:matchedXsect}
\end{eqnarray}
where $\mathcal{N}_{ij}$ contains the flux factor and the average over
spins and colours of the initial state partons of flavour $i$ and $j$,
e.g.~$\mathcal{N}_{gg} = \frac{1}{2s} \frac{1}{64} \frac{1}{4}$. 
And we have introduced the luminosity factors $\mathcal{L}_{ij}$,
defined by
\begin{align}
\label{eq:Xsection}
\sigma(S) &= \int_{\tau_{min}}^{1} \mathrm{d} \tau \, 
\sum_{ij} \int_{\tau}^{1} \frac{\mathrm{d} x}{x} f_{i} (x,\mu_F) \, f_{j} (\frac{\tau}{x},\mu_F) \, 
\sigma_{ij}(s = \tau S)\, \nonumber\\
&\equiv \sum_{ij} \int_{\tau_{min}}^{1} \mathrm{d} \tau \frac{\mathrm{d} \mathcal{L}_{ij}}{\mathrm{d} \tau} \, 
\sigma_{ij}(s = \tau S),
\end{align}
where $f_{i}(x,\mu_F)$ is the parton distribution function (PDF) of a parton with momentum fraction $x$ and flavour
$i$ (including gluons) and $\mu_F$ is the factorisation scale.
The 2- and 3-particle phase-space 
integration measures are denoted by $\mathrm{d}\Phi_2$ and $\mathrm{d}\Phi_3$.
The symbol $\mathbf{c} \equiv 32\,\!\pi \, \alpha_e \, Q_t^2 \, T_R\,\delta^{a_1a_2}$  collects constants which have been extracted in the
definition of ${\mathrm M}_t$.
The real-radiation contributions with the factors of $a_s$ extracted are symbolically denoted by ${\cal M}_{R,[ij]}$ 
and the collinear-subtraction counterterm is denoted by $\sigma_C$.
We do not include resummation effects in the real-radiation because it is suppressed by a factor of
$\beta$.
The ${\mathrm M}_{\rm{OC}}^{(0)}$ and ${\mathrm M}_{\rm{OC}}^{(1)}(\mu)$ denote the LO and NLO double-counted part 
of the amplitude as we discussed above. Note that the explicit dependence of ${\mathrm M}_{\rm{OC}}^{(1)}(\mu)$ on the scale $\mu$ 
stems from the renormalisation of the Green's function $G(\vec0;{ E})$, while $\mu_R$ comes from the renormalisation of UV divergences 
in ${\cal  M}_V(\mu_R)$ and $\mu_F$ from initial-state collinear factorisation.

For the numerical evaluation of eq.~\eqref{eq:matchedXsect}, we expand ${\mathrm M}^{(0)}_{\mathrm{OC}}$ and ${\mathrm M}^{(1)}_{\mathrm{OC}}$
to respectively $\mathcal{O}(\beta^3)$ and $\mathcal{O}(\beta^2)$ using the expressions stated in Section~\ref{sec:pnrqcd_amp}. 
At the two-loop order, the UV-renormalised and IR-subtracted fixed-order amplitude ${\mathrm M}_{t}$ has a Coulomb
singularity which is logarithmically divergent in the limit $\beta\rightarrow{}0$. This singularity is, however,
subtracted by the expanded term ${\mathrm M}_{\mathrm{OC}}$, while a resummed description of the Coulomb interactions 
is added back by the term ${\mathcal B}_{t} \, G(\vec0;{\mathcal E})$.
For this purpose, we evaluate the Schr\"odinger equation~\eqref{eq:schroedinger} numerically~\cite{Kiyo:2010jm} to obtain $G(\vec0;{\mathcal E})$,
where we include $\mathcal{O}(\als)$ corrections to the QCD potential~\cite{Fischler:1977yf,Billoire:1979ih}. Unlike the calculation in~\cite{Kawabata:2016aya}, we also
include $\mathcal{O}(\als)$ corrections to ${\mathcal B}_t$ as listed above.

\section{Results}
\label{sec:results}

Our numerical results are calculated at a hadronic centre-of-mass energy of $13$\,TeV, using the parton distribution functions
PDF4LHC15{\tt\_}nlo{\tt\_}100~\cite{Butterworth:2015oua,CT14,MMHT14,NNPDF}
interfaced  via
LHAPDF~\cite{Buckley:2014ana}, along with the corresponding value for
$\alpha_s$.  For the
electromagnetic coupling, we use $\alpha = 1/137.035999139$.
The mass of the top quark is fixed to $m_t=173$\,GeV. The top-quark width is set to zero in
the fixed order calculation, and to $\Gamma_t=1.498$\,GeV in the numerical
evaluation of the Green's function $G(\vec0;{\mathcal E},\mu)$ in accordance with Ref.~\cite{Kawabata:2016aya}.
We use the cuts $p_{T,\gamma_1}^{\rm{min}}=40$\,GeV,
$p_{T,\gamma_2}^{\rm{min}}=25$\,GeV and $|\eta_\gamma|\leq 2.5$.
No photon isolation cuts are applied.

The factorisation and renormalisation scale uncertainties are
estimated by varying the scales $\mu_{F}$ and $\mu_{R}$. Unless
specified otherwise, the scale variation bands 
represent the envelopes of a 7-point scale variation with
$\mu_{R,F}=c_{R,F}\,m_{\gamma\gamma}/2$, where $c_R,c_F\in \{2,1,0.5\}$ and where the
extreme variations $(c_R,c_F)=(2,0.5)$ and $(c_R,c_F)=(0.5,2)$ have
been omitted. The dependence on the scale $\mu$ introduced by renormalisation of the Green's
function $G(\vec{r};{\mathcal E})$ in our NRQCD matched results is investigated separately.

\subsection{Validation}

\subsection*{Fixed-order calculation}

We have validated the massless NLO cross section by comparison to
MCFM version 9.0~\cite{Campbell:2019dru,Campbell:2011bn} and find agreement within the
numerical uncertainties for all scale choices. We also compared against the
results shown in~\cite{Maltoni:2018zvp} and found agreement for the
central scale choice, however we found a smaller scale uncertainty band
than in the originally published version of Ref.~\cite{Maltoni:2018zvp}. The authors of
Ref.~\cite{Maltoni:2018zvp} meanwhile have sent us an updated version
of their figures, where we find agreement.

We remark that the helicity amplitudes can also be computed via first performing the Lorentz tensor decomposition, using the form factor projectors given in Ref.~\cite{Binoth:2002xg}, and then evaluating contractions between the corresponding Lorentz structures and external polarisation vectors in 4 dimensions using the spinor-helicity representations. This amounts to obtaining helicity amplitudes defined in the t'Hooft-Veltman scheme~\cite{tHooft:1972tcz}. 
We confirm numerically that the same finite remainders are obtained for all helicity configurations at a few chosen test points (while the unsubtracted helicity  amplitudes do differ starting from the subleading power in $\epsilon$).

As a further cross check, we evaluate our amplitude with $t \leftrightarrow u$ interchanged and confirm that the helicity amplitudes are permuted as expected.

\begin{figure}[hptb]
    \includegraphics[width=\textwidth]{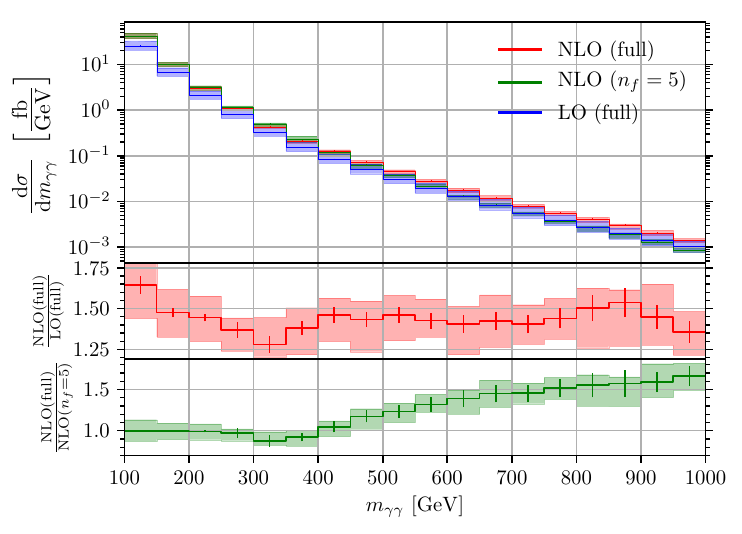}
    \caption{Diphoton invariant mass distribution (fixed order calculation), comparing the result with $n_f=5$ to the result including massive top quark loops.
             The lower panels show the ratios
             $\mathrm{NLO(full)}/\mathrm{LO(full)}$ and $\mathrm{NLO(full)}/\mathrm{NLO}(n_f=5)$. 
             The shaded bands show the envelope of the 7-point scale variation as explained in the text.
             In the ratio plots, only the scale of the numerators is varied, while the scale of the denominators is  fixed  to $\mu_{R}=\mu_{F}=\mgg/2$.
             The bars indicate the uncertainty due to the numerical
             evaluation of the phase-space and loop integrals.}
    \label{fig:myy}
\end{figure}

\subsection*{NRQCD amplitude}

Numerical values for the coefficients ${\mathcal A}^{J}_{t,\{\lambda_i\}}$  at leading-order in $\alpha_s$ up to $J=4$ are given in  Ref.~\cite{Kawabata:2016aya}.
We have used them as a check of our numerical calculation of the Born amplitude.

We also evaluated the massive two-loop amplitude at $615$ phase space points with $m_t =
173$\,GeV in the ranges $0 < \cos{}\left(\theta{}\right) < 1$ and
$0.001 \leq{} \beta{} \leq{} 0.2$, using the program \pysecdec{}~\cite{Borowka:2017idc,Borowka:2018goh}.
The amplitude can numerically be fitted to a suitable ansatz in $\beta$ and $\cos\theta$.
We have compared the coefficients of terms proportional to $\ln{(\beta)}$ to the known analytical results based on expanding equation~\eqref{eq:2loopexpansion}
and find good agreement. Note that the coefficients of terms not proportional to $\ln{(\beta)}$ receive contributions from the unknown term $A^{(1)}(\theta)$
and can therefore not be checked this way.

\subsection{Invariant mass distribution of the diphoton system}

The distribution of the invariant mass of the photon pair is shown in
Fig.~\ref{fig:myy} for invariant masses up to 1\,TeV, where we show
purely fixed order results at LO, at NLO with five massless flavours
and at NLO including massive top quark loops. The ratio plots show the
K-factor including the full quark loop content and the ratio between the full
and the five-flavour NLO cross-section.
We observe that the scale uncertainties are reduced at NLO, and that
the top quark loops enhance the differential cross section for $\mgg$
values far beyond the top-quark pair-production threshold, asymptotically
approaching the $n_f=6$ value~\cite{Campbell:2016yrh}.

\begin{figure}[hptb]
    \includegraphics[width=\textwidth]{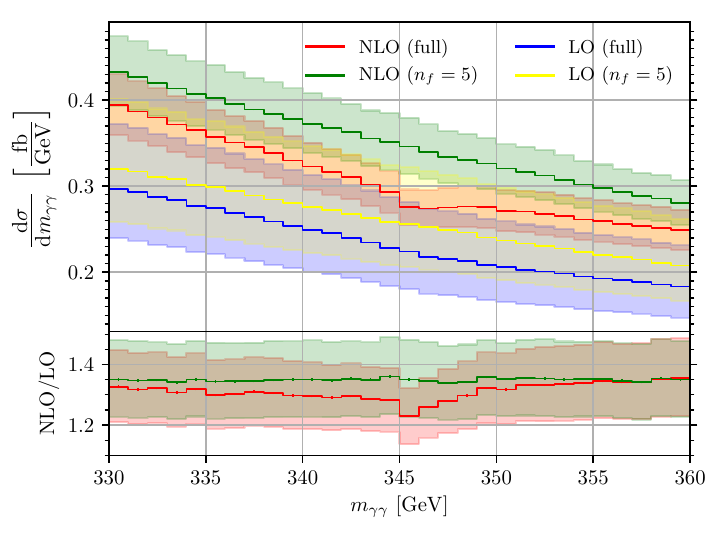}
    \caption{Zoom into the threshold region of the diphoton invariant mass distribution (fixed order calculation), showing the $n_f=5$ and full result separately. The shaded
            bands indicate the scale uncertainties, while the bars indicate uncertainties due to the numerical
            evaluation of the phase-space and loop integrals. The ratio plot in the lower panel shows the ratios $\mathrm{NLO(full)}/\mathrm{LO(full)}$ (red) and
            $\mathrm{NLO(}n_f=5\mathrm{)}/\mathrm{LO(}n_f=5\mathrm{)}$
          (green), with the scale variation bands obtained by varying the scale in the numerators only.}
    \label{fig:myy_toponly_zoom}
\end{figure}

In Fig.~\ref{fig:myy_toponly_zoom} we zoom into the threshold region,
still showing fixed order results only. We can clearly see that after
the top quark pair production threshold, the full result shows a dip
and then changes
slope, which is due to the fact that the two-loop amplitude contains
the exchange of a Coulomb gluon (see top left diagram of Fig.~\ref{fig:2l_diagrams}),
as explained in Section~\ref{sec:threshold}.
In Ref.~\cite{Kawabata:2016aya} it was suggested that this
characteristic ``dip-bump structure'' could be used for a
determination of the top quark mass which is free from top quark
reconstruction uncertainties, at least at the FCC where the
statistical uncertainties for this process would be very small, and
the systematic uncertainty due to the finite resolution of the photon
energies and angles should be at least as good as at the LHC, where it
is at the sub-percent level~\cite{Aad:2014nim,Khachatryan:2015iwa}.

\begin{figure}[hptb]
    \includegraphics[width=\textwidth]{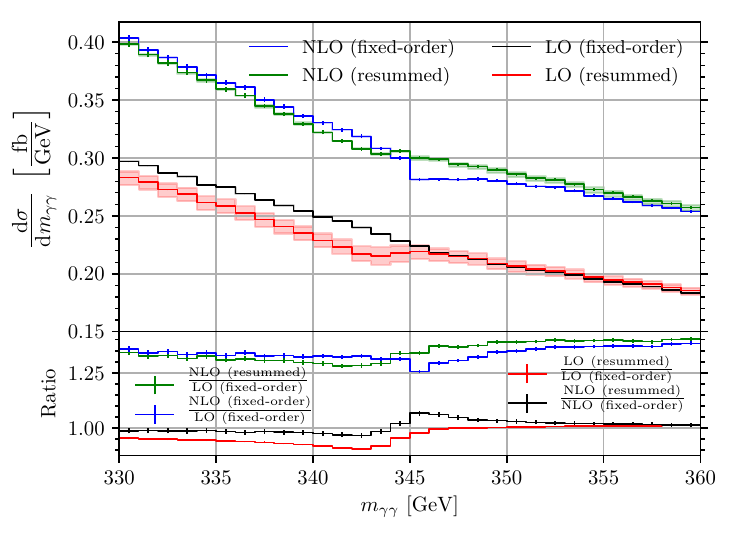}
    \caption{Zoom into the threshold region of the diphoton invariant mass distribution, comparing results with and without NRQCD.
             The shaded bands indicate the scale uncertainty by varying the scale $\mu$ by a factor of 2 around the central scale $\mu=80$\,GeV.
             The renormalisation and the factorisation scales are set to $\mu_R=\mu_F=m_{\gamma\gamma}/2$ and not varied in this plot.
             The bars indicate uncertainties due to the numerical evaluation of the phase-space and loop integrals.}
    \label{fig:myy_NRQCD}
  \end{figure}

In Fig.~\ref{fig:myy_NRQCD} we show the $\mgg$-distribution in the
threshold region which results from 
a combination of the fixed-order NLO (QCD) calculation with the resummation of Coulomb gluon exchanges
as described in Section~\ref{sec:pnrqcd_Xsect}. The scale
band in this figure are produced by varying only $\mu$, the scale
associated to the renormalisation of the Green's function.
We observe that the dependence on the scale $\mu$
is considerably reduced at NLO compared to the leading-order matched
cross-section. The scale band at NLO is comparable to the size of our
numerical uncertainties.
Further, our leading-order matched cross-section shows a milder dependence on $\mu$ than the one presented in~\cite{Kawabata:2016aya}.
This is due to the inclusion of NLO-terms in the coefficient ${\mathcal B}_{t}(\beta)$, which have been omitted
in~\cite{Kawabata:2016aya}.

We do not consider the effects from a colour-octet $t\bar{t}$ state
because the corresponding Green's function is monotonically increasing
in the resonance region~\cite{Kiyo:2008bv} and therefore not expected to
move the position of the dip significantly.

Now let us address the prospects to measure the top quark mass from
the threshold behaviour of the $\mgg$ distribution.
In Ref.~\cite{Kawabata:2016aya} it was argued that the characteristic
dip-bump structure does not change its location in the $\mgg$
spectrum under scale variations, only the overall normalisation is
changing.
It was also
anticipated that the inclusion of the fixed order two-loop amplitude
would reduce this uncertainty. Indeed we find that the NLO corrections
reduce the scale uncertainties due to 7-point $\mu_R,
\mu_F$-variations from about 20\% at LO to just below the 10\% level
at NLO.

\begin{figure}[hptb]
  \includegraphics[width=\textwidth]{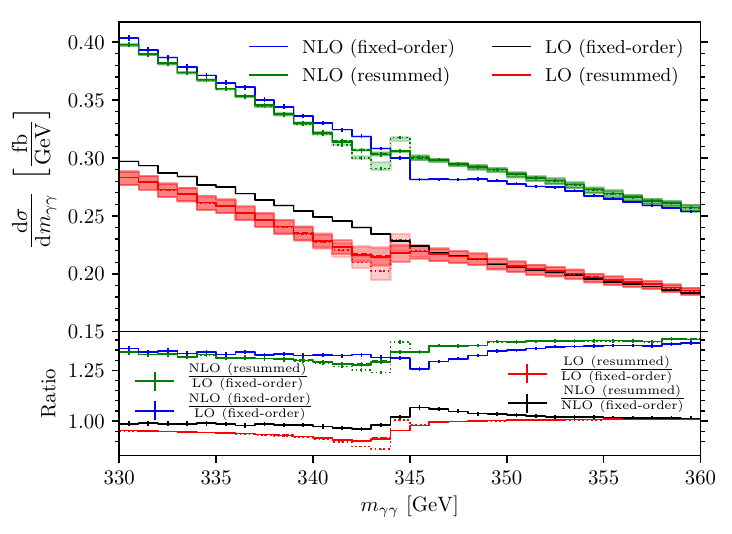}
    \caption{Effect of the top quark width on the dip-bump-structure
      at LO and NLO. The solid lines correspond to $\Gamma_t=  1.367$\,GeV, the dashed to $\Gamma_t= 1.498$\,GeV
and the dotted to $\Gamma_t=0.5$\,GeV. The corresponding scale bands are also shown in the upper plot.}
    \label{fig:topwidth}
  \end{figure}
  
The treatment of the top quark width
included in the NRQCD calculation could be
refined by including higher order corrections to the width.
We have  investigated how a change of the width
affects the height and the location of the dip-bump-structure.
We have performed the calculation with three different values for $\Gamma_t$:  our default LO
value of $\Gamma_t=1.498$\,GeV, an NLO value of $\Gamma_t=1.367$\,GeV, obtained using
the expressions of Ref.~\cite{Jezabek:1988iv}, and an ``extreme" value
of $\Gamma_t=0.5$\,GeV.
The result is shown in Fig.~\ref{fig:topwidth}.
We observe that the amplitude of the dip-bump-structure is quite sensitive
to the width, with small widths giving a larger
dip-bump-amplitude. 
This feature might offer the possibility to
constrain the top quark width based on a template fit to the $\mgg$
distribution, similar to what has been performed in Ref.~\cite{ATLAS:2019onj} for
the $m_{lb}$ distribution.
Furthermore, we found that at LO, the dip-bump-structure is less
broad for smaller top quark widths, while with 1\,GeV binnings this is not visible at NLO. 

Our results show that the description of the region which is critical
for the top quark mass measurement sensitively depends on the
theoretical modelling. Therefore, without calculating even higher
orders, it is diffcult to assess how large the uncertainties due to
the theoretical description really are.

The experimental resolution at the LHC is estimated to be about
$10\%/\sqrt{E_\gamma\,\rm{[GeV]}}$~\cite{Aad:2014nim,Khachatryan:2015iwa}.
A resolution of the photon energy scale of about 0.5\% or
better leads to a systematic uncertainty on $m_t$ of about
1\,GeV~\cite{Kawabata:2016aya}.
Such an uncertainty is not competitive with current measurements from
top quark pair production~\cite{Castro:2019ttg}. Therefore such a
top quark mass determination has to wait for measurements at a future
collider if at all feasible.
In order to assess from the theory side whether the shape change present in our best
theoretical prediction would be sufficient to measure the top quark
mass, it would be useful to perform the current study for various top
quark masses, which would enter an experimental template fit.
However such an analysis is beyond the scope of this paper and we postpone
it to future work.

\section{Conclusions and Outlook}

\label{sec:conclusions}

We have calculated the production of a photon pair in gluon fusion at
order $\alpha_s^3$, including massive top quark loops.
This calculation, which is NLO for the gluon initiated channel, is
formally part of the N$^3$LO corrections to the $pp\to\gamma\gamma$
process. However, the gluon channel is important at the LHC due to the
large gluon luminosity.
The top quark loops have a considerable impact on the diphoton
invariant mass spectrum, at values of $\mgg$ larger than about 800\,GeV
they enhance the $\mgg$
differential cross section by more than 50\%.

The region around the top quark pair production threshold in the
diphoton invariant mass spectrum is particularly interesting.
The fixed order amplitude has a  divergence starting at two loops due
to Coulomb gluon exchange. We have used NRQCD methods to resum the bound
state effects in order to obtain a more reliable description of the
threshold region. Matching the resummed calculation to our fixed order NLO
calculation we observe a reduction of the renormalisation and
factorisation scale uncertainties in the
threshold region by more than a factor of two, and an even more
drastic reduction of the scale uncertainty related to the renormalised
NLO Green's function.

These results are promising in view of the possibility of measuring the
top quark mass from the characteristic behaviour of the diphoton
invariant mass spectrum  around the top quark pair production
threshold.
In Ref.~\cite{Kawabata:2016aya}, it was found that the LO resummed result shows a characteristic ``dip-bump'' structure
and the conclusion was that this would allow a precise measurement of
the top quark mass with the statistics and photon resolution
projected for the FCC, once an NLO calculation is available such that
the scale uncertainties are reduced. Now we indeed found that at NLO, the scale
uncertainties are reduced. Furthermore, the 
characteristic ``dip-bump'' structure at NLO remains stable when switching
from a LO value to an NLO value for the top quark width.
A detailed assessment of whether this structure and the change in slope
is pronounced enough for a top quark mass measurement
once all channels contributing to this observable
are included deserves further study. It also requires a detailed study of  the prospective
experimental uncertainties.

Furthermore, it would be interesting to investigate other top quark
mass schemes, as well as the possibility to constrain the top quark
width from this process.
However  this is beyond the scope of this paper and therefore we defer it to future work.

\section*{Acknowledgements}
We would like to thank Matteo Becchetti and Roberto Bonciani for
providing integrals in analytic form (which we finally did not use).
We also thank Manoj Mandal, Xiaoran Zhao, Michael Spira and Robert
Szafron for interesting discussions and Fabio Maltoni for conversation about the scale uncertainties.
This research was supported in part by the COST Action CA16201 (`Particleface') of the European Union,
and by the Swiss National Science Foundation (SNF) under grant number 200020-175595.
The research of JS was supported by the European Union through the ERC Advanced Grant MC@NNLO (340983).

\bibliographystyle{JHEP}
\bibliography{main_ggyy.bib}

\providecommand{\href}[2]{#2}\begingroup\raggedright\begin{thebibliography}{10}

\bibitem{Chatrchyan:2014fsa}
{\scshape CMS} collaboration, S.~Chatrchyan et~al., \emph{{Measurement of
  differential cross sections for the production of a pair of isolated photons
  in pp collisions at $\sqrt{s}=7\,\text {TeV} $}},
  \href{https://doi.org/10.1140/epjc/s10052-014-3129-3}{\emph{Eur. Phys. J.}
  {\bfseries C74} (2014) 3129}
  [\href{https://arxiv.org/abs/1405.7225}{{\ttfamily 1405.7225}}].

\bibitem{Aaboud:2017vol}
{\scshape ATLAS} collaboration, M.~Aaboud et~al., \emph{{Measurements of
  integrated and differential cross sections for isolated photon pair
  production in $pp$ collisions at $\sqrt{s}=8$ TeV with the ATLAS detector}},
  \href{https://doi.org/10.1103/PhysRevD.95.112005}{\emph{Phys. Rev.}
  {\bfseries D95} (2017) 112005}
  [\href{https://arxiv.org/abs/1704.03839}{{\ttfamily 1704.03839}}].

\bibitem{Dicus:1987fk}
D.~A. Dicus and S.~S.~D. Willenbrock, \emph{{Photon Pair Production and the
  Intermediate Mass Higgs Boson}},
  \href{https://doi.org/10.1103/PhysRevD.37.1801}{\emph{Phys. Rev.} {\bfseries
  D37} (1988) 1801}.

\bibitem{Dixon:2003yb}
L.~J. Dixon and M.~S. Siu, \emph{{Resonance continuum interference in the
  diphoton Higgs signal at the LHC}},
  \href{https://doi.org/10.1103/PhysRevLett.90.252001}{\emph{Phys. Rev. Lett.}
  {\bfseries 90} (2003) 252001}
  [\href{https://arxiv.org/abs/hep-ph/0302233}{{\ttfamily hep-ph/0302233}}].

\bibitem{Martin:2012xc}
S.~P. Martin, \emph{{Shift in the LHC Higgs diphoton mass peak from
  interference with background}},
  \href{https://doi.org/10.1103/PhysRevD.86.073016}{\emph{Phys. Rev.}
  {\bfseries D86} (2012) 073016}
  [\href{https://arxiv.org/abs/1208.1533}{{\ttfamily 1208.1533}}].

\bibitem{deFlorian:2013psa}
D.~de~Florian, N.~Fidanza, R.~J. Hernandez-Pinto, J.~Mazzitelli,
  Y.~Rotstein~Habarnau and G.~F.~R. Sborlini, \emph{{A complete $O(\alpha_S^2)$
  calculation of the signal-background interference for the Higgs diphoton
  decay channel}},
  \href{https://doi.org/10.1140/epjc/s10052-013-2387-9}{\emph{Eur. Phys. J.}
  {\bfseries C73} (2013) 2387}
  [\href{https://arxiv.org/abs/1303.1397}{{\ttfamily 1303.1397}}].

\bibitem{Martin:2013ula}
S.~P. Martin, \emph{{Interference of Higgs diphoton signal and background in
  production with a jet at the LHC}},
  \href{https://doi.org/10.1103/PhysRevD.88.013004}{\emph{Phys. Rev.}
  {\bfseries D88} (2013) 013004}
  [\href{https://arxiv.org/abs/1303.3342}{{\ttfamily 1303.3342}}].

\bibitem{Dixon:2013haa}
L.~J. Dixon and Y.~Li, \emph{{Bounding the Higgs Boson Width Through
  Interferometry}},
  \href{https://doi.org/10.1103/PhysRevLett.111.111802}{\emph{Phys. Rev. Lett.}
  {\bfseries 111} (2013) 111802}
  [\href{https://arxiv.org/abs/1305.3854}{{\ttfamily 1305.3854}}].

\bibitem{Campbell:2017rke}
J.~Campbell, M.~Carena, R.~Harnik and Z.~Liu, \emph{{Interference in the
  $gg\rightarrow h \rightarrow \gamma\gamma$ On-Shell Rate and the Higgs Boson
  Total Width}}, \href{https://doi.org/10.1103/PhysRevLett.119.199901,
  10.1103/PhysRevLett.119.181801}{\emph{Phys. Rev. Lett.} {\bfseries 119}
  (2017) 181801} [\href{https://arxiv.org/abs/1704.08259}{{\ttfamily
  1704.08259}}].

\bibitem{Cieri:2017kpq}
L.~Cieri, F.~Coradeschi, D.~de~Florian and N.~Fidanza,
  \emph{{Transverse-momentum resummation for the signal-background interference
  in the $H\to\gamma\gamma$ channel at the LHC}},
  \href{https://doi.org/10.1103/PhysRevD.96.054003}{\emph{Phys. Rev.}
  {\bfseries D96} (2017) 054003}
  [\href{https://arxiv.org/abs/1706.07331}{{\ttfamily 1706.07331}}].

\bibitem{Aaboud:2016tru}
{\scshape ATLAS} collaboration, M.~Aaboud et~al., \emph{{Search for resonances
  in diphoton events at $\sqrt{s}$=13 TeV with the ATLAS detector}},
  \href{https://doi.org/10.1007/JHEP09(2016)001}{\emph{JHEP} {\bfseries 09}
  (2016) 001} [\href{https://arxiv.org/abs/1606.03833}{{\ttfamily
  1606.03833}}].

\bibitem{Sirunyan:2018wnk}
{\scshape CMS} collaboration, A.~M. Sirunyan et~al., \emph{{Search for physics
  beyond the standard model in high-mass diphoton events from proton-proton
  collisions at $\sqrt{s} =$ 13 TeV}},
  \href{https://doi.org/10.1103/PhysRevD.98.092001}{\emph{Phys. Rev.}
  {\bfseries D98} (2018) 092001}
  [\href{https://arxiv.org/abs/1809.00327}{{\ttfamily 1809.00327}}].

\bibitem{Jain:2016kai}
S.~R. Dugad, P.~Jain, S.~Mitra, P.~Sanyal and R.~K. Verma, \emph{{The top
  threshold effect in the $\gamma\gamma$ production at the LHC}},
  \href{https://doi.org/10.1140/epjc/s10052-018-6188-z}{\emph{Eur. Phys. J.}
  {\bfseries C78} (2018) 715}
  [\href{https://arxiv.org/abs/1605.07360}{{\ttfamily 1605.07360}}].

\bibitem{Kawabata:2016aya}
S.~Kawabata and H.~Yokoya, \emph{{Top-quark mass from the diphoton mass
  spectrum}}, \href{https://doi.org/10.1140/epjc/s10052-017-4884-8}{\emph{Eur.
  Phys. J.} {\bfseries C77} (2017) 323}
  [\href{https://arxiv.org/abs/1607.00990}{{\ttfamily 1607.00990}}].

\bibitem{Binoth:1999qq}
T.~Binoth, J.~P. Guillet, E.~Pilon and M.~Werlen, \emph{{A Full next-to-leading
  order study of direct photon pair production in hadronic collisions}},
  \href{https://doi.org/10.1007/s100520050024}{\emph{Eur. Phys. J.} {\bfseries
  C16} (2000) 311} [\href{https://arxiv.org/abs/hep-ph/9911340}{{\ttfamily
  hep-ph/9911340}}].

\bibitem{Balazs:2007hr}
C.~Balazs, E.~L. Berger, P.~M. Nadolsky and C.~P. Yuan, \emph{{Calculation of
  prompt diphoton production cross-sections at Tevatron and LHC energies}},
  \href{https://doi.org/10.1103/PhysRevD.76.013009}{\emph{Phys. Rev.}
  {\bfseries D76} (2007) 013009}
  [\href{https://arxiv.org/abs/0704.0001}{{\ttfamily 0704.0001}}].

\bibitem{Bern:2001df}
Z.~Bern, A.~De~Freitas and L.~J. Dixon, \emph{{Two loop amplitudes for gluon
  fusion into two photons}},
  \href{https://doi.org/10.1088/1126-6708/2001/09/037}{\emph{JHEP} {\bfseries
  09} (2001) 037} [\href{https://arxiv.org/abs/hep-ph/0109078}{{\ttfamily
  hep-ph/0109078}}].

\bibitem{Bern:2002jx}
Z.~Bern, L.~J. Dixon and C.~Schmidt, \emph{{Isolating a light Higgs boson from
  the diphoton background at the CERN LHC}},
  \href{https://doi.org/10.1103/PhysRevD.66.074018}{\emph{Phys. Rev.}
  {\bfseries D66} (2002) 074018}
  [\href{https://arxiv.org/abs/hep-ph/0206194}{{\ttfamily hep-ph/0206194}}].

\bibitem{Campbell:2011bn}
J.~M. Campbell, R.~K. Ellis and C.~Williams, \emph{{Vector boson pair
  production at the LHC}},
  \href{https://doi.org/10.1007/JHEP07(2011)018}{\emph{JHEP} {\bfseries 07}
  (2011) 018} [\href{https://arxiv.org/abs/1105.0020}{{\ttfamily 1105.0020}}].

\bibitem{Maltoni:2018zvp}
F.~Maltoni, M.~K. Mandal and X.~Zhao, \emph{{Top-quark effects in diphoton
  production through gluon fusion at NLO in QCD}},
  \href{https://arxiv.org/abs/1812.08703}{{\ttfamily 1812.08703}}.

\bibitem{Czakon:2008zk}
M.~Czakon, \emph{{Tops from Light Quarks: Full Mass Dependence at Two-Loops in
  QCD}}, \href{https://doi.org/10.1016/j.physletb.2008.05.028}{\emph{Phys.
  Lett.} {\bfseries B664} (2008) 307}
  [\href{https://arxiv.org/abs/0803.1400}{{\ttfamily 0803.1400}}].

\bibitem{Mandal:2018cdj}
M.~K. Mandal and X.~Zhao, \emph{{Evaluating multi-loop Feynman integrals
  numerically through differential equations}},
  \href{https://doi.org/10.1007/JHEP03(2019)190}{\emph{JHEP} {\bfseries 03}
  (2019) 190} [\href{https://arxiv.org/abs/1812.03060}{{\ttfamily
  1812.03060}}].

\bibitem{Caron-Huot:2014lda}
S.~Caron-Huot and J.~M. Henn, \emph{{Iterative structure of finite loop
  integrals}}, \href{https://doi.org/10.1007/JHEP06(2014)114}{\emph{JHEP}
  {\bfseries 06} (2014) 114} [\href{https://arxiv.org/abs/1404.2922}{{\ttfamily
  1404.2922}}].

\bibitem{Becchetti:2017abb}
M.~Becchetti and R.~Bonciani, \emph{{Two-Loop Master Integrals for the Planar
  QCD Massive Corrections to Di-photon and Di-jet Hadro-production}},
  \href{https://doi.org/10.1007/JHEP01(2018)048}{\emph{JHEP} {\bfseries 01}
  (2018) 048} [\href{https://arxiv.org/abs/1712.02537}{{\ttfamily
  1712.02537}}].

\bibitem{vonManteuffel:2017hms}
A.~von Manteuffel and L.~Tancredi, \emph{{A non-planar two-loop three-point
  function beyond multiple polylogarithms}},
  \href{https://doi.org/10.1007/JHEP06(2017)127}{\emph{JHEP} {\bfseries 06}
  (2017) 127} [\href{https://arxiv.org/abs/1701.05905}{{\ttfamily
  1701.05905}}].

\bibitem{Broedel:2019hyg}
J.~Broedel, C.~Duhr, F.~Dulat, B.~Penante and L.~Tancredi, \emph{{Elliptic
  polylogarithms and Feynman parameter integrals}},
  \href{https://doi.org/10.1007/JHEP05(2019)120}{\emph{JHEP} {\bfseries 05}
  (2019) 120} [\href{https://arxiv.org/abs/1902.09971}{{\ttfamily
  1902.09971}}].

\bibitem{Aglietti:2006tp}
U.~Aglietti, R.~Bonciani, G.~Degrassi and A.~Vicini, \emph{{Analytic Results
  for Virtual QCD Corrections to Higgs Production and Decay}},
  \href{https://doi.org/10.1088/1126-6708/2007/01/021}{\emph{JHEP} {\bfseries
  01} (2007) 021} [\href{https://arxiv.org/abs/hep-ph/0611266}{{\ttfamily
  hep-ph/0611266}}].

\bibitem{Anastasiou:2006hc}
C.~Anastasiou, S.~Beerli, S.~Bucherer, A.~Daleo and Z.~Kunszt, \emph{{Two-loop
  amplitudes and master integrals for the production of a Higgs boson via a
  massive quark and a scalar-quark loop}},
  \href{https://doi.org/10.1088/1126-6708/2007/01/082}{\emph{JHEP} {\bfseries
  01} (2007) 082} [\href{https://arxiv.org/abs/hep-ph/0611236}{{\ttfamily
  hep-ph/0611236}}].

\bibitem{Catani:2011qz}
S.~Catani, L.~Cieri, D.~de~Florian, G.~Ferrera and M.~Grazzini, \emph{{Diphoton
  production at hadron colliders: a fully-differential QCD calculation at
  NNLO}}, \href{https://doi.org/10.1103/PhysRevLett.108.072001,
  10.1103/PhysRevLett.117.089901}{\emph{Phys. Rev. Lett.} {\bfseries 108}
  (2012) 072001} [\href{https://arxiv.org/abs/1110.2375}{{\ttfamily
  1110.2375}}].

\bibitem{Catani:2018krb}
S.~Catani, L.~Cieri, D.~de~Florian, G.~Ferrera and M.~Grazzini, \emph{{Diphoton
  production at the LHC: a QCD study up to NNLO}},
  \href{https://doi.org/10.1007/JHEP04(2018)142}{\emph{JHEP} {\bfseries 04}
  (2018) 142} [\href{https://arxiv.org/abs/1802.02095}{{\ttfamily
  1802.02095}}].

\bibitem{Campbell:2016yrh}
J.~M. Campbell, R.~K. Ellis, Y.~Li and C.~Williams, \emph{{Predictions for
  diphoton production at the LHC through NNLO in QCD}},
  \href{https://doi.org/10.1007/JHEP07(2016)148}{\emph{JHEP} {\bfseries 07}
  (2016) 148} [\href{https://arxiv.org/abs/1603.02663}{{\ttfamily
  1603.02663}}].

\bibitem{Grazzini:2017mhc}
M.~Grazzini, S.~Kallweit and M.~Wiesemann, \emph{{Fully differential NNLO
  computations with MATRIX}},
  \href{https://doi.org/10.1140/epjc/s10052-018-5771-7}{\emph{Eur. Phys. J.}
  {\bfseries C78} (2018) 537}
  [\href{https://arxiv.org/abs/1711.06631}{{\ttfamily 1711.06631}}].

\bibitem{Chen:2019wyb}
L.~Chen, \emph{{A prescription for projectors to compute helicity amplitudes in
  D dimensions}},  \href{https://arxiv.org/abs/1904.00705}{{\ttfamily
  1904.00705}}.

\bibitem{Ahmed:2019udm}
T.~Ahmed, A.~H. Ajjath, L.~Chen, P.~K. Dhani, P.~Mukherjee and V.~Ravindran,
  \emph{{Polarised Amplitudes and Soft-Virtual Cross Sections for $b\bar b
  \rightarrow ZH$ at NNLO in QCD}},
  \href{https://arxiv.org/abs/1910.06347}{{\ttfamily 1910.06347}}.

\bibitem{Karplus:1950zz}
R.~Karplus and M.~Neuman, \emph{{The scattering of light by light}},
  \href{https://doi.org/10.1103/PhysRev.83.776}{\emph{Phys. Rev.} {\bfseries
  83} (1951) 776}.

\bibitem{Bern:2001dg}
Z.~Bern, A.~De~Freitas, L.~J. Dixon, A.~Ghinculov and H.~L. Wong, \emph{{QCD
  and QED corrections to light by light scattering}},
  \href{https://doi.org/10.1088/1126-6708/2001/11/031}{\emph{JHEP} {\bfseries
  11} (2001) 031} [\href{https://arxiv.org/abs/hep-ph/0109079}{{\ttfamily
  hep-ph/0109079}}].

\bibitem{Binoth:2002xg}
T.~Binoth, E.~W.~N. Glover, P.~Marquard and J.~J. van~der Bij, \emph{{Two loop
  corrections to light by light scattering in supersymmetric QED}},
  \href{https://doi.org/10.1088/1126-6708/2002/05/060}{\emph{JHEP} {\bfseries
  05} (2002) 060} [\href{https://arxiv.org/abs/hep-ph/0202266}{{\ttfamily
  hep-ph/0202266}}].

\bibitem{Bern:1995db}
Z.~Bern and A.~G. Morgan, \emph{{Massive loop amplitudes from unitarity}},
  \href{https://doi.org/10.1016/0550-3213(96)00078-8}{\emph{Nucl. Phys.}
  {\bfseries B467} (1996) 479}
  [\href{https://arxiv.org/abs/hep-ph/9511336}{{\ttfamily hep-ph/9511336}}].

\bibitem{Bernicot:2008th}
C.~Bernicot, \emph{{Light-light amplitude from generalized unitarity in massive
  QED}},  \href{https://arxiv.org/abs/0804.0749}{{\ttfamily 0804.0749}}.

\bibitem{Frixione:1995ms}
S.~Frixione, Z.~Kunszt and A.~Signer, \emph{{Three jet cross-sections to
  next-to-leading order}},
  \href{https://doi.org/10.1016/0550-3213(96)00110-1}{\emph{Nucl. Phys.}
  {\bfseries B467} (1996) 399}
  [\href{https://arxiv.org/abs/hep-ph/9512328}{{\ttfamily hep-ph/9512328}}].

\bibitem{Nason:2004rx}
P.~Nason, \emph{{A New method for combining NLO QCD with shower Monte Carlo
  algorithms}},
  \href{https://doi.org/10.1088/1126-6708/2004/11/040}{\emph{JHEP} {\bfseries
  11} (2004) 040} [\href{https://arxiv.org/abs/hep-ph/0409146}{{\ttfamily
  hep-ph/0409146}}].

\bibitem{Frixione:2007vw}
S.~Frixione, P.~Nason and C.~Oleari, \emph{{Matching NLO QCD computations with
  Parton Shower simulations: the POWHEG method}},
  \href{https://doi.org/10.1088/1126-6708/2007/11/070}{\emph{JHEP} {\bfseries
  11} (2007) 070} [\href{https://arxiv.org/abs/0709.2092}{{\ttfamily
  0709.2092}}].

\bibitem{Alioli:2010xd}
S.~Alioli, P.~Nason, C.~Oleari and E.~Re, \emph{{A general framework for
  implementing NLO calculations in shower Monte Carlo programs: the POWHEG
  BOX}}, \href{https://doi.org/10.1007/JHEP06(2010)043}{\emph{JHEP} {\bfseries
  06} (2010) 043} [\href{https://arxiv.org/abs/1002.2581}{{\ttfamily
  1002.2581}}].

\bibitem{Jones:2016bci}
S.~P. Jones, \emph{{Automation of 2-loop Amplitude Calculations}},
  \href{https://doi.org/10.22323/1.260.0069}{\emph{PoS} {\bfseries LL2016}
  (2016) 069} [\href{https://arxiv.org/abs/1608.03846}{{\ttfamily
  1608.03846}}].

\bibitem{vonManteuffel:2012np}
A.~von Manteuffel and C.~Studerus, \emph{{Reduze 2 - Distributed Feynman
  Integral Reduction}},  \href{https://arxiv.org/abs/1201.4330}{{\ttfamily
  1201.4330}}.

\bibitem{Argeri:2014qva}
M.~Argeri, S.~Di~Vita, P.~Mastrolia, E.~Mirabella, J.~Schlenk, U.~Schubert
  et~al., \emph{{Magnus and Dyson Series for Master Integrals}},
  \href{https://doi.org/10.1007/JHEP03(2014)082}{\emph{JHEP} {\bfseries 03}
  (2014) 082} [\href{https://arxiv.org/abs/1401.2979}{{\ttfamily 1401.2979}}].

\bibitem{vonManteuffel:2014qoa}
A.~von Manteuffel, E.~Panzer and R.~M. Schabinger, \emph{{A quasi-finite basis
  for multi-loop Feynman integrals}},
  \href{https://doi.org/10.1007/JHEP02(2015)120}{\emph{JHEP} {\bfseries 02}
  (2015) 120} [\href{https://arxiv.org/abs/1411.7392}{{\ttfamily 1411.7392}}].

\bibitem{Borowka:2017idc}
S.~Borowka, G.~Heinrich, S.~Jahn, S.~P. Jones, M.~Kerner, J.~Schlenk et~al.,
  \emph{{pySecDec: a toolbox for the numerical evaluation of multi-scale
  integrals}}, \href{https://doi.org/10.1016/j.cpc.2017.09.015}{\emph{Comput.
  Phys. Commun.} {\bfseries 222} (2018) 313}
  [\href{https://arxiv.org/abs/1703.09692}{{\ttfamily 1703.09692}}].

\bibitem{Borowka:2018goh}
S.~Borowka, G.~Heinrich, S.~Jahn, S.~P. Jones, M.~Kerner and J.~Schlenk,
  \emph{{A GPU compatible quasi-Monte Carlo integrator interfaced to
  pySecDec}}, \href{https://doi.org/10.1016/j.cpc.2019.02.015}{\emph{Comput.
  Phys. Commun.} {\bfseries 240} (2019) 120}
  [\href{https://arxiv.org/abs/1811.11720}{{\ttfamily 1811.11720}}].

\bibitem{Luisoni:2013cuh}
G.~Luisoni, P.~Nason, C.~Oleari and F.~Tramontano, \emph{{$HW^{\pm}$/HZ + 0 and
  1 jet at NLO with the POWHEG BOX interfaced to GoSam and their merging within
  MiNLO}}, \href{https://doi.org/10.1007/JHEP10(2013)083}{\emph{JHEP}
  {\bfseries 1310} (2013) 083}
  [\href{https://arxiv.org/abs/1306.2542}{{\ttfamily 1306.2542}}].

\bibitem{Cullen:2011ac}
G.~Cullen, N.~Greiner, G.~Heinrich, G.~Luisoni, P.~Mastrolia, G.~Ossola et~al.,
  \emph{{Automated One-Loop Calculations with GoSam}},
  \href{https://doi.org/10.1140/epjc/s10052-012-1889-1}{\emph{Eur. Phys. J.}
  {\bfseries C72} (2012) 1889}
  [\href{https://arxiv.org/abs/1111.2034}{{\ttfamily 1111.2034}}].

\bibitem{Cullen:2014yla}
G.~Cullen et~al., \emph{{G$\scriptsize{O}$S$\scriptsize{AM}$-2.0: a tool for
  automated one-loop calculations within the Standard Model and beyond}},
  \href{https://doi.org/10.1140/epjc/s10052-014-3001-5}{\emph{Eur. Phys. J.}
  {\bfseries C74} (2014) 3001}
  [\href{https://arxiv.org/abs/1404.7096}{{\ttfamily 1404.7096}}].

\bibitem{Caswell:1985ui}
W.~E. Caswell and G.~P. Lepage, \emph{{Effective Lagrangians for Bound State
  Problems in QED, QCD, and Other Field Theories}},
  \href{https://doi.org/10.1016/0370-2693(86)91297-9}{\emph{Phys. Lett.}
  {\bfseries 167B} (1986) 437}.

\bibitem{Bodwin:1994jh}
G.~T. Bodwin, E.~Braaten and G.~P. Lepage, \emph{{Rigorous QCD analysis of
  inclusive annihilation and production of heavy quarkonium}},
  \href{https://doi.org/10.1103/PhysRevD.55.5853,
  10.1103/PhysRevD.51.1125}{\emph{Phys. Rev.} {\bfseries D51} (1995) 1125}
  [\href{https://arxiv.org/abs/hep-ph/9407339}{{\ttfamily hep-ph/9407339}}].

\bibitem{Pineda:1997bj}
A.~Pineda and J.~Soto, \emph{{Effective field theory for ultrasoft momenta in
  NRQCD and NRQED}},
  \href{https://doi.org/10.1016/S0920-5632(97)01102-X}{\emph{Nucl. Phys. Proc.
  Suppl.} {\bfseries 64} (1998) 428}
  [\href{https://arxiv.org/abs/hep-ph/9707481}{{\ttfamily hep-ph/9707481}}].

\bibitem{Beneke:1997zp}
M.~Beneke and V.~A. Smirnov, \emph{{Asymptotic expansion of Feynman integrals
  near threshold}},
  \href{https://doi.org/10.1016/S0550-3213(98)00138-2}{\emph{Nucl. Phys.}
  {\bfseries B522} (1998) 321}
  [\href{https://arxiv.org/abs/hep-ph/9711391}{{\ttfamily hep-ph/9711391}}].

\bibitem{Melnikov:1994jb}
K.~Melnikov, M.~Spira and O.~I. Yakovlev, \emph{{Threshold effects in two
  photon decays of Higgs particles}},
  \href{https://doi.org/10.1007/BF01560100}{\emph{Z. Phys.} {\bfseries C64}
  (1994) 401} [\href{https://arxiv.org/abs/hep-ph/9405301}{{\ttfamily
  hep-ph/9405301}}].

\bibitem{Melnikov:1993np}
K.~Melnikov and O.~I. Yakovlev, \emph{{Top near threshold: All alpha-S
  corrections are trivial}},
  \href{https://doi.org/10.1016/0370-2693(94)90410-3}{\emph{Phys. Lett.}
  {\bfseries B324} (1994) 217}
  [\href{https://arxiv.org/abs/hep-ph/9302311}{{\ttfamily hep-ph/9302311}}].

\bibitem{Fischler:1977yf}
W.~Fischler, \emph{{Quark - anti-Quark Potential in QCD}},
  \href{https://doi.org/10.1016/0550-3213(77)90026-8}{\emph{Nucl. Phys.}
  {\bfseries B129} (1977) 157}.

\bibitem{Billoire:1979ih}
A.~Billoire, \emph{{How Heavy Must Be Quarks in Order to Build Coulombic q
  anti-q Bound States}},
  \href{https://doi.org/10.1016/0370-2693(80)90279-8}{\emph{Phys. Lett.}
  {\bfseries 92B} (1980) 343}.

\bibitem{Beneke:1998rk}
M.~Beneke, \emph{{A Quark mass definition adequate for threshold problems}},
  \href{https://doi.org/10.1016/S0370-2693(98)00741-2}{\emph{Phys. Lett.}
  {\bfseries B434} (1998) 115}
  [\href{https://arxiv.org/abs/hep-ph/9804241}{{\ttfamily hep-ph/9804241}}].

\bibitem{Hoang:1998xf}
A.~H. Hoang and T.~Teubner, \emph{{Top quark pair production at threshold:
  Complete next-to-next-to-leading order relativistic corrections}},
  \href{https://doi.org/10.1103/PhysRevD.58.114023}{\emph{Phys. Rev.}
  {\bfseries D58} (1998) 114023}
  [\href{https://arxiv.org/abs/hep-ph/9801397}{{\ttfamily hep-ph/9801397}}].

\bibitem{Beneke:1999ff}
M.~Beneke, A.~Signer and V.~A. Smirnov, \emph{{A Two loop application of the
  threshold expansion: The Bottom quark mass from b anti-b production}},  in
  \emph{{Radiative corrections: Application of quantum field theory to
  phenomenology. Proceedings, 4th International Symposium, RADCOR'98,
  Barcelona, Spain, September 8-12, 1998}}, pp.~223--234, 1999,
  \href{https://arxiv.org/abs/hep-ph/9906476}{{\ttfamily hep-ph/9906476}}.

\bibitem{Hoang:2001mm}
A.~H. Hoang, A.~V. Manohar, I.~W. Stewart and T.~Teubner, \emph{{The Threshold
  t anti-t cross-section at NNLL order}},
  \href{https://doi.org/10.1103/PhysRevD.65.014014}{\emph{Phys. Rev.}
  {\bfseries D65} (2002) 014014}
  [\href{https://arxiv.org/abs/hep-ph/0107144}{{\ttfamily hep-ph/0107144}}].

\bibitem{Petrelli:1997ge}
A.~Petrelli, M.~Cacciari, M.~Greco, F.~Maltoni and M.~L. Mangano, \emph{{NLO
  production and decay of quarkonium}},
  \href{https://doi.org/10.1016/S0550-3213(97)00801-8}{\emph{Nucl. Phys.}
  {\bfseries B514} (1998) 245}
  [\href{https://arxiv.org/abs/hep-ph/9707223}{{\ttfamily hep-ph/9707223}}].

\bibitem{Hagiwara:2008df}
K.~Hagiwara, Y.~Sumino and H.~Yokoya, \emph{{Bound-state Effects on Top Quark
  Production at Hadron Colliders}},
  \href{https://doi.org/10.1016/j.physletb.2008.07.006}{\emph{Phys. Lett.}
  {\bfseries B666} (2008) 71}
  [\href{https://arxiv.org/abs/0804.1014}{{\ttfamily 0804.1014}}].

\bibitem{Kiyo:2008bv}
{Kiyo, Y. and K\"uhn, Johann H. and Moch, S. and Steinhauser, M. and Uwer, P.},
  \emph{{Top-quark pair production near threshold at LHC}},
  \href{https://doi.org/10.1140/epjc/s10052-009-0892-7}{\emph{Eur. Phys. J.}
  {\bfseries C60} (2009) 375}
  [\href{https://arxiv.org/abs/0812.0919}{{\ttfamily 0812.0919}}].

\bibitem{Hoang:2004tg}
A.~H. Hoang and C.~J. Reisser, \emph{{Electroweak absorptive parts in NRQCD
  matching conditions}},
  \href{https://doi.org/10.1103/PhysRevD.71.074022}{\emph{Phys. Rev.}
  {\bfseries D71} (2005) 074022}
  [\href{https://arxiv.org/abs/hep-ph/0412258}{{\ttfamily hep-ph/0412258}}].

\bibitem{Kiyo:2010jm}
Y.~Kiyo, A.~Pineda and A.~Signer, \emph{{New determination of inclusive
  electromagnetic decay ratios of heavy quarkonium from QCD}},
  \href{https://doi.org/10.1016/j.nuclphysb.2010.08.007}{\emph{Nucl. Phys.}
  {\bfseries B841} (2010) 231}
  [\href{https://arxiv.org/abs/1006.2685}{{\ttfamily 1006.2685}}].

\bibitem{Butterworth:2015oua}
J.~Butterworth et~al., \emph{{PDF4LHC recommendations for LHC Run II}},
  \href{https://arxiv.org/abs/1510.03865}{{\ttfamily 1510.03865}}.

\bibitem{CT14}
S.~Dulat, T.-J. Hou, J.~Gao, M.~Guzzi, J.~Huston, P.~Nadolsky et~al.,
  \emph{{New parton distribution functions from a global analysis of quantum
  chromodynamics}},
  \href{https://doi.org/10.1103/PhysRevD.93.033006}{\emph{Phys. Rev.}
  {\bfseries D93} (2016) 033006}
  [\href{https://arxiv.org/abs/1506.07443}{{\ttfamily 1506.07443}}].

\bibitem{MMHT14}
L.~A. Harland-Lang, A.~D. Martin, P.~Motylinski and R.~S. Thorne, \emph{{Parton
  distributions in the LHC era: MMHT 2014 PDFs}},
  \href{https://doi.org/10.1140/epjc/s10052-015-3397-6}{\emph{Eur. Phys. J.}
  {\bfseries C75} (2015) 204}
  [\href{https://arxiv.org/abs/1412.3989}{{\ttfamily 1412.3989}}].

\bibitem{NNPDF}
{\scshape NNPDF} collaboration, R.~D. Ball et~al., \emph{{Parton distributions
  for the LHC Run II}},
  \href{https://doi.org/10.1007/JHEP04(2015)040}{\emph{JHEP} {\bfseries 04}
  (2015) 040} [\href{https://arxiv.org/abs/1410.8849}{{\ttfamily 1410.8849}}].

\bibitem{Buckley:2014ana}
A.~Buckley, J.~Ferrando, S.~Lloyd, K.~Nordstr{\"o}m, B.~Page, M.~R{\"u}fenacht
  et~al., \emph{{LHAPDF6: parton density access in the LHC precision era}},
  \href{https://doi.org/10.1140/epjc/s10052-015-3318-8}{\emph{Eur. Phys. J.}
  {\bfseries C75} (2015) 132}
  [\href{https://arxiv.org/abs/1412.7420}{{\ttfamily 1412.7420}}].

\bibitem{Campbell:2019dru}
J.~Campbell and T.~Neumann, \emph{{Precision Phenomenology with MCFM}},
  \href{https://arxiv.org/abs/1909.09117}{{\ttfamily 1909.09117}}.

\bibitem{tHooft:1972tcz}
G.~'t~Hooft and M.~J.~G. Veltman, \emph{{Regularization and Renormalization of
  Gauge Fields}},
  \href{https://doi.org/10.1016/0550-3213(72)90279-9}{\emph{Nucl. Phys.}
  {\bfseries B44} (1972) 189}.

\bibitem{Aad:2014nim}
{\scshape ATLAS} collaboration, G.~Aad et~al., \emph{{Electron and photon
  energy calibration with the ATLAS detector using LHC Run 1 data}},
  \href{https://doi.org/10.1140/epjc/s10052-014-3071-4}{\emph{Eur. Phys. J.}
  {\bfseries C74} (2014) 3071}
  [\href{https://arxiv.org/abs/1407.5063}{{\ttfamily 1407.5063}}].

\bibitem{Khachatryan:2015iwa}
{\scshape CMS} collaboration, V.~Khachatryan et~al., \emph{{Performance of
  Photon Reconstruction and Identification with the CMS Detector in
  Proton-Proton Collisions at sqrt(s) = 8 TeV}},
  \href{https://doi.org/10.1088/1748-0221/10/08/P08010}{\emph{JINST} {\bfseries
  10} (2015) P08010} [\href{https://arxiv.org/abs/1502.02702}{{\ttfamily
  1502.02702}}].

\bibitem{Jezabek:1988iv}
M.~Jezabek and J.~H. Kuhn, \emph{{QCD Corrections to Semileptonic Decays of
  Heavy Quarks}},
  \href{https://doi.org/10.1016/0550-3213(89)90108-9}{\emph{Nucl. Phys.}
  {\bfseries B314} (1989) 1}.

\bibitem{ATLAS:2019onj}
{The ATLAS collaboration}, \emph{{Measurement of the top-quark decay width in
  top-quark pair events in the dilepton channel at $\sqrt{s}=13$ TeV with the
  ATLAS detector}},
  \href{https://arxiv.org/abs/ATLAS-CONF-2019-038}{{\ttfamily
  ATLAS-CONF-2019-038}}.

\bibitem{Castro:2019ttg}
{\scshape ATLAS, CMS} collaboration, A.~Castro, \emph{{Top Quark Mass
  Measurements in ATLAS and CMS}},  in \emph{{12th International Workshop on
  Top Quark Physics (TOP2019) Beijing, China, September 22-27, 2019}}, 2019,
  \href{https://arxiv.org/abs/1911.09437}{{\ttfamily 1911.09437}}.

\end{thebibliography}\endgroup

\end{document}